\documentstyle[12pt, epsfig]{article}
\newcommand{\lsim}{\raisebox{-0.13cm}{~\shortstack{$<$ \\[-0.07cm] $\sim$}}~}
\newcommand{\gsim}{\raisebox{-0.13cm}{~\shortstack{$>$ \\[-0.07cm] $\sim$}}~}

\newcommand{\defprod}{\raisebox{-0.13cm}{~\shortstack{$\prod$ \\[-0.07cm] 
${}_{j\neq k}$}}~}

\def\cc#1{\kern .7em\hfill #1 \hfill\kern .7em}
\begin{document}
\baselineskip 18pt
\begin{flushright}
PM/01-45 \\
hep-ph/0112353
\end{flushright}
\begin{center}
{\Large \bf Infrared Quasi Fixed Point Structure \\ in Extended Yukawa Sectors\\
and\\
\vspace{.2cm}
Application to R-parity Violation}

\vspace{1cm}

{\sc  Y. Mambrini$^a$,  G. Moultaka$^b$ } 

\vspace{0.5cm}  {\it $^a$, Laboratoire de Physique Th\'eorique \\ 
Universit\'e Paris XI, Batiment 210, F-91405 Orsay Cedex, France}

\vspace{0.3cm}
{\it  $^b$ Physique Math\'ematique et Th\'eorique, UMR No 5825--CNRS, \\
Universit\'e Montpellier II, F--34095 Montpellier Cedex 5, France.
}
\end{center}
\vspace{1cm}

\begin{abstract}
\noindent
We investigate one-loop renormalization group evolutions of extended 
sectors of Yukawa type couplings. It is shown that Landau Poles  which usually 
provide necessary low energy upper bounds that saturate quickly with 
increasing initial value conditions, lead in some cases to the opposite 
behaviour: some of the low energy couplings decrease and become 
vanishingly small for increasingly large initial conditions! 
We write down the general criteria for this to happen in typical situations, 
highlighting a concept of {\sl repulsive} quasi-fixed points, 
and illustrate the case both within a two-Yukawa toy model as well as in the 
minimal supersymmetric standard model with R-parity violation. 
In the latter case we consider the theoretical upper bounds on the various 
couplings, identifying regimes where $\lambda_{kl3}, \lambda'_{kkk}, 
\lambda''_{3kl}$ are dynamically suppressed due to the Landau Pole. 
We stress the importance of considering a large number 
of couplings simultaneously. This leads altogether to a phenomenologically 
interesting seesaw effect in the magnitudes of the various R-parity violating 
couplings, complementing and in some cases improving the existing limits.         
\end{abstract}

\newpage

\section{Introduction}
The Infrared Quasi Fixed Point (IRQFP) structure in the Yukawa sector, 
\cite{hill}, has played an important role in singling out the supersymmetric 
extension of the standard model in the small $\tan \beta$ regime, as a natural 
framework to accommodate a top quark mass of $\sim 175$ GeV \cite{wagneretal}, 
as opposed to the standard model 
itself or to alternatives such as the gauged 
Nambu--Jona-Lasinio \cite{BHL} model, which predicted altogether too heavy a
top quark ($\gsim 200$ GeV). The experimental (almost) negative search of
 Higgs particle pushes it's mass above 114 GeV in the  standard model scenario
\cite{LEPHiggs1, LEPHiggs3}, or above 89 GeV \cite{LEPHiggs2, LEPHiggs3}  
 in the Minimal Supersymmetric Standard Model (MSSM), ruling out the small 
$\tan \beta$ IRQFP regime.  This motivates naturally the study of 
supersymmetric scenarios where non zero bottom and $\tau$ quark Yukawa 
couplings are of the same order of magnitude as that of the top  
\cite{largetbeta}. By the same token, 
one can also consider  further extensions such as the next to minimal 
supersymmetric standard model ((M+1)SSM) \cite{NMSSM}, or the MSSM without 
R-parity \cite{RPV}, leading to an increasing number of Yukawa type couplings.\\

The aim of the present paper is to study the generic behaviour of the
renormalization group evolutions of the Yukawa couplings in the presence
of a large number of these couplings, and in particular the effect of Landau 
Poles on the perturbative bounds. The main result is that for increasingly large
values of the initial conditions, some of the couplings can be 
suppressed (formally to zero!) at low energies, a behaviour which differs 
drastically from the usually expected one \cite{hill}. 
This phenomenon is related 
to the appearance of a rich structure of Infrared Quasi Fixed Points, some of 
which can have an unconventional {\sl repulsive} character. 
In practice this means that the close numerical connection between the Landau 
Pole and the perturbativity bound, usually observed for the top quark Yukawa 
coupling for instance, does not prevail for other couplings in more general configurations.\\

A natural place where this new mechanism occurs is the MSSM with R-parity 
violation ($\not{\!\!R}_p$-MSSM). Here one has 45 Yukawa couplings as new free 
parameters to be added to those of the MSSM. In the absence of theoretically 
compelling reasons to put them to zero (apart from the elegant \cite{fayet}, 
yet {\sl ad hoc}, R-parity!) the new parameters have to be allowed 
for and studied phenomenologically for their own sake. They can influence many 
physical processes and have already been experimentally constrained this way 
\cite{expRPV}. An important motivation for constraining R-parity violating 
scenarios is the bearing it can have on the nature of the supersymmetric dark 
matter candidate in the Universe, and indirectly on the 
origin of supersymmetry breaking. While conserved R-parity stabilizes the 
lightest supersymmetric particle (LSP) and tends to favour the lightest 
neutralino, rather than the gravitino, as the LSP (a natural configuration in
gravity mediated supersymmetry breaking scenarios), even an extremely small 
violation of R-parity becomes a "Damocles sword" over the neutralino LSP as a
serious candidate for cold dark matter. Conversely, a moderate amount of 
R-parity violation would favour the gravitino as a potential dark matter 
candidate \cite{gravitino}, and to some extent favour scenarios with low scale 
supersymmetry breaking.\\


\noindent
The model-independent mechanism we elucidate in this paper
 can have interesting implications on the study of R-parity violation.
The rest of the paper will be thus divided into two main parts  
--the first part, corresponding to sections {\bf 2} and {\bf 3} and appendices 
{\bf A} and {\bf B}, is devoted to a detailed discussion of the generic 
solutions of the one-loop renormalization group equations for an arbitrary 
number of Yukawa couplings, and in particular the
dynamical attraction to null couplings due to the presence of Landau Poles 
--the second part, {\sl i.e.} section {\bf 4} and 
appendix {\bf C},  is an illustration of these features in the context of 
$\not{\!\!R}_p$-MSSM. There we study the effect of increasing the number of
active couplings by considering subsets with 6, 9 and 13 non-zero couplings.
We compare our results with previous theoretical bounds as well as with 
experimental limits. In appendix {\bf C} we write explicitly the one-loop
renormalization group equations governing the evolution of a large
set of couplings. In section {\bf 5} we conclude and summarize our results. 
[The reader interested mainly in phenomenological 
implications on R-parity violation  can go directly to  section {\bf 4} which 
is fairly self-contained.]

\section{General Formulation}
\subsection{Integrated Forms for the Running Yukawa Couplings}

Let us consider first a general gauge theory with an arbitrary number of gauge 
group factors and  of fermion multiplets coupled to some scalar 
fields. Through this paper we will always assume no flavour violating 
fermion-fermion-scalar Yukawa couplings. The renormalization group
equations (RGE) for the gauge couplings, and [with some further simplifying 
assumptions about the scalar self couplings] those for the Yukawa couplings, 
take the following form at the one-loop level  \cite{chengetal},\\

\begin{eqnarray}
\frac{d}{dt} \tilde{\alpha}_i&=&-b_i \tilde{\alpha}^2_i \label{gieq} \\
\frac{d}{dt} \tilde{Y}_k(t) &=&\tilde{Y}_k(t)(\sum_ic_{ki} \tilde{\alpha}_i(t)-\sum_la_{kl}\tilde{Y}_l(t)) 
\label{yukeq}
\end{eqnarray}

\noindent
where $t \equiv Log[ M_{GUT}^2/Q^2]$ denotes the scale evolution parameter, 
and we define 

\begin{equation}
\tilde{Y}_k \equiv h_k^2/(16\pi^2),  \;\;\tilde{\alpha}_i 
\equiv g^2_i/(16\pi^2) \label{defeq}
\end{equation}

where $h_k, g_i$ denote respectively the Yukawa and gauge couplings.  
The $c_{ki}$ and $a_{kl}$ are constant coefficients depending on the model.

\noindent
The general solution for such a system (valid for an arbitrary number of Yukawa
couplings labeled by $k$) reads \cite{auberson}
\begin{eqnarray}
\tilde{\alpha}_i(t)&=&\frac{\tilde{\alpha}_i(0)}{1+b_i \tilde{\alpha}_i(0) t} \nonumber \\
\tilde{Y}_k(t)&=&\frac{\tilde{Y}_k(0) u_k(t)}{1+a_{kk}\tilde{Y}_k(0)\int_0^tu_k(t')dt'} 
\label{Ysol}
\end{eqnarray}

\noindent
where the auxiliary functions $u_k$ are given by

\begin{equation}
u_k(t)= \frac{ E_k(t)}{\defprod ( 1 + a_{jj} \tilde{Y}_j(0) \int_0^t u_j(t')dt')^{a_{kj}/a_{jj}} }
\label{usol}
\end{equation}

\noindent
and the functions $E_k(t) \equiv exp[ \int_0^t \sum_ic_{ki} \tilde{\alpha}_i(t')dt']$
read explicitly 

\begin{equation}
E_k(t)=\prod \limits_{i=1}^3(1+b_i \tilde{\alpha}_i(0)t)^{\frac{c_{ki}}{b_i}}
\label{Ek}
\end{equation}

\noindent
In \cite{auberson} we called these solutions ``integrated forms" to stress
the fact that they are not explicit but rather iterative. As
it turned out, and will be illustrated once more in this paper, they
allow, even in this form, a generic extraction of several interesting features,
which could be hardly pinned down from mere numerics, in particular in relation with the infrared quasi fixed points. 
In many physically interesting cases such as the non 
supersymmetric standard model, or the MSSM or the NMSSM, the RGE's for the 
Yukawa couplings  take indeed the form of Eq.(\ref{yukeq}) and the above 
solutions are directly applicable. This is no more strictly true in the general 
case of $\not{\!\!R}_p$-MSSM, nonetheless, we will see that these solutions 
still prove very efficient in describing this case too.  
  
\subsection{Asymptotic behaviour and quasi fixed points}

Avoiding Landau poles in the Yukawa system leads to consistency upper
bounds on the values of the running Yukawa  couplings at some
low energy scale. These bounds can be most straightforwardly
determined by looking at the asymptotic behaviour
of the running couplings when some, or all, of the initial
conditions are taken infinitely large. We will come back
at length to the meaning of these bounds in the subsequent sections. Here we 
derive for further reference the complete procedure which allows a non ambiguous
determination of the asymptotic behaviour, recalling and extending the
point made in ref. \cite{comment}. It is useful to define

\begin{equation}
r_{kj} = \frac{\tilde{Y}_k(0)}{\tilde{Y}_j(0)} \label{rat}
\end{equation}

\noindent
and study the asymptotic behaviour by increasing the $\tilde{Y}_i(0)$'s while 
keeping the above scaling ratios fixed and finite ($\neq 0$). One can thus 
carry out the discussion in terms of the set of finite $r$'s and one single 
initial condition parameter  $Y(0) \equiv Y^0$ which is allowed to become 
infinite.  In this limit the values taken by the $\tilde{Y}_k(t)$ at a low 
energy scale $t$ are the so-called infrared effective fixed points or 
Infrared  Quasi Fixed Points (IRQFP)\footnote{
Let us mention briefly here a different definition 
adopted in \cite{bargeretal}, where one requires 
$\frac{d}{d\tau} \tilde{Y}_k(\tau) |_{\tau=t} \simeq 0$ for all $k$, at
a given low scale $t$ and solves the resulting linear system of equations in 
the {\sl positive} $\tilde{Y}_k(t)$'s from Eqs.(\ref{yukeq}). There is,
 however, no guarantee in general that this linear system has positive 
solutions; one then has to change accordingly the amount of ``zeroing"
of $\frac{d}{d\tau} \tilde{Y}_k(\tau)$ or/and the scale $t$. Our definition
is more general in that the slowing down of the running is automatically such that the IRQFP's always exist}. From Eq.(\ref{usol}) one expects
the $u$'s to behave like

\begin{equation}
u_k^\infty \equiv \frac{u_k^{\mathrm{QFP}}(r_{ij})}{{(Y^0)}^{p_k}} 
\label{uinfty}
\end{equation}

\noindent
for asymptotically large $Y^0$, where we will refer to the $p_k$'s as 
``asymptotic powers". Here $u_k^{\mathrm{QFP}}$ is an integrated form which
is $Y^0$ independent but may or may not depend on the scaling parameters $r$. 
Now it is important to stress that Eq.(\ref{uinfty}) is quite general and 
{\sl does not} mean\footnote{With that 
respect an unfortunate erroneous statement slipped in refs. \cite{pandita, pandita1}. } that we neglect the $1$ in the denominator of Eq.(\ref{usol}) due to 
increasingly large $\tilde{Y}_j(0)$'s. It is indeed easy to see from 
Eqs.(\ref{Ysol}, \ref{uinfty}) that the IRQFP are controlled by the various 
asymptotic powers in the following way,

\begin{eqnarray}
\tilde{Y}_k^{\mathrm{QFP}}(t)&=&\frac{u_k^{\mathrm{QFP}}(t)}{a_{kk} 
\int_0^t u_k^{\mathrm{QFP}}(t')dt'} \; \;\;\;\; \;\;\; \mbox{if } \; p_k <1 \nonumber \\
\tilde{Y}_k^{\mathrm{QFP}}(t)&=&\frac{u_k^{\mathrm{QFP}}(t)}{1 + a_{kk} 
\int_0^t u_k^{\mathrm{QFP}}(t')dt'} \; \; \mbox{if } \; p_k =1 \label{YQFP} \\
\tilde{Y}_k^{\mathrm{QFP}}(t)&=& 0 ~~~~~~~~~~~~~~~~~~~~~ \;\;\;\;\;\;\; \; \mbox{if } \; p_k>1 \nonumber 
\label{YsolFP}
\end{eqnarray}

\noindent
Only when $p_k <1$ is it justified to drop $1$ in the denominator. In this case
the IRQFP's take the usual form \cite{hill} as it was shown for the MSSM
\cite{kazakov}. The two other cases in Eq.(\ref{YQFP}) occur for
larger Yukawa sectors as it was found for the (M+1)SSM in \cite{mambrini} and
for the $\not{\!\!R}_p$-MSSM in \cite{comment}. In particular, the very
unusual configuration where some IRQFP's vanish is worth attention. As was 
stressed in \cite{comment}, the prior determination of the asymptotic powers is 
crucial. Straightforward inspection shows that they must be solutions  
of the following equation,
\begin{equation}
\vec{ {\cal P}} = {\cal M} \cdot ( 1 - \vec{ {\cal P}}) 
\theta( 1 - \vec{ {\cal P}}) \vec{\delta}
\label{matricegen}
\end{equation}

\noindent
Here $\vec{{\cal P}}$ is a column vector of all the asymptotic powers $p_k$;
$( 1 - \vec{ {\cal P}}) \theta( 1 - \vec{ {\cal P}})$ is a shorthand for
the column vector with components $(1 - p_k) \theta(1  - p_k) \delta_k$ where 
$\theta$ is the Heaviside function, $\delta_k=1 \; \mbox{or} \; 0$ 
respectively for $\tilde{Y}_k(0)$ infinite or finite; ${\cal M}$ is defined by
 
\begin{equation}
({\cal M})_{kj} = \frac{a_{kj}}{a_{jj}} \label{matricedef}
\end{equation}

\noindent
Note that due to the $\theta$ function in Eq.(\ref{matricegen}) 
this system is linear in the $p_k$'s only in patches. It can thus allow
simultaneously for more than one set of solutions which should
correspond to alternated configurations given by Eq.(\ref{YQFP}). We will
study at length  the meaning of such multiple solutions in the following
sections, highlighting the (new) phenomenon of {\sl repulsive} IRQFP. 

\subsection{An example: the two-Yukawa system}

\noindent
It is instructive to illustrate the above as well as the forthcoming features
of the paper in the simplest case of two Yukawa couplings 
(the ``top/bottom" system):

\begin{eqnarray}
\frac{d}{dt} \tilde{Y}_1(t) &=&\tilde{Y}_1(t)(\sum_ic_{1i} g^2_i(t)
- a_{11}\tilde{Y}_1(t) - a_{12}\tilde{Y}_2(t)) \nonumber \\
\frac{d}{dt} \tilde{Y}_2(t) &=&\tilde{Y}_2(t)(\sum_ic_{2i} g^2_i(t)
- a_{21}\tilde{Y}_1(t) - a_{22}\tilde{Y}_2(t)) 
\label{ytoy}
\end{eqnarray}

\noindent
where generically $a_{12} = a_{21} \equiv |a|, a_{11} = a_{22} \equiv |b|$
as in the MSSM (or in the Standard Model (SM)). We will stick to this 
configuration for simplicity although our discussion applies more generally. 
In this case Eq.(\ref{matricegen}) reads

\begin{equation}
\left(
\begin{array}{l}
p_1 \cr
p_2 \cr
\end{array}
\right)
=
\left(
\begin{array}{cc}
0 & \frac{|a|}{|b|} \cr
\frac{|a|}{|b|} & 0 \cr
\end{array}
\right)
\left(
\begin{array}{l}
(1 - p_1) \; \theta[1 - p_1]  \cr
(1 - p_2) \; \theta[1 - p_2]  \cr
\end{array}
\right)
\label{matricetoy}
\end{equation}

\noindent
The critical issue here is whether $\frac{|a|}{|b|} < $ or $> 1$.
As can be seen from Appendix B.1, in the first case Eq.(\ref{matricetoy})
has a unique solution and thus one single Infrared Quasi Fixed Point
given by the first equation in (\ref{YQFP}).
This case corresponds to realistic models
like the SM and the MSSM (where $\frac{|a|}{|b|} = 1/3$ and $1/6$ respectively).
In the ``toy model" case $\frac{|a|}{|b|} >1$,  
comparing Eqs.(\ref{matricetoy}) and (\ref{matrice1}) 
({\sl i.e.} $\alpha=\delta=0$ and $\beta=\gamma= |a|/|b|$), one finds
(see  Eqs.(\ref{crit1}, \ref{crit2})) that 
 there are three different solutions to Eq.(\ref{matricetoy}) corresponding
to three possible IRQFP's with the appropriate configurations in Eq.(\ref{YQFP}).
Which one of these solutions is {\sl dynamically} chosen by the system
will be addressed in the next section. This will serve as a useful 
illustration, in a simple setting,  
of a phenomenon which actually occurs in realistic models with a large number
of Yukawa couplings, such as R-parity violating models to be discussed 
in section {\bf 4}. 

\section{Multiple IRQFPs and Landau Pole free domains}

Let us first state the two main points we will be discussing 
in the next two subsections:
 
\begin{itemize}
\item[{\bf (I)}] The IR Quasi Fixed Points discussed in the previous section, 
and studied in the literature, 
are only isolated points on the boundary of the Landau Pole free domains  (LPfd)
which will be defined by Eqs.(\ref{landaupole}) (or by Eqs.(\ref{landaupole1})).
Thus the ``rectangle domains" given by the Quasi Fixed Points
encode part {\sl but not all} of the constraints on the allowed low scale 
values of the Yukawa couplings.     
\item[{\bf (II)}] In extended Yukawa sectors and depending on the ratios
$a_{kj}/a_{jj}$ there can appear simultaneously several IR Quasi Fixed 
Points for the very same configuration of increasingly large initial conditions!
Actually, some of these points are {\sl repulsive} and others {\sl attractive},
depending on the way one approaches the given configuration
of initial conditions. \\

\end{itemize}

\noindent
The relevance of point {\bf (II)} will become manifest through the specific
examples of the next sections. We just anticipate here two important and
mutually related consequences: 
{\sl (i)} the attractive IRQFP's lead to a dynamical suppression
of some of the Yukawa couplings and {\sl (ii)} the perturbativity bound
  becomes, numerically, substantially different  from the Landau Pole bound. 
Furthermore, it should  be clear that the multiple solution scenario we will 
be discussing is not to be confused with the fact that one obtains
different quasi fixed points for different sets of finite/infinite Yukawa 
initial conditions 
(like for the typical examples of  ``small $\tan \beta$" or ``large  
$\tan \beta$" fixed point scenarios \cite{wagneretal, largetbeta}).

\subsection{LPfd beyond the rectangular approximation}

Hereafter we determine the equations defining the  Landau Pole
free domains (LPfd) beyond the crude ``rectangular approximation".
It is convenient for the discussion to write Eq.(\ref{Ysol}) at two distinct 
scales $t$ and $t^0 (< t)$ (thus $t$ corresponds to an energy scale
{\sl lower} than that of $t^0$). 
Eliminating $\tilde{Y}_k(0)$ in the ensuing equations, one obtains the 
general correlations between the Yukawa couplings at two arbitrary scales,
expressed in a bottom-up approach where the Yukawa couplings at the high
scale $t^0$ are cast in terms of their initial conditions at the low scale $t$ ,

\begin{eqnarray}
\tilde{Y}_k(t^0)&=&\frac{\tilde{Y}_k(t) u_k(t^0, t)}{1- a_{kk} \tilde{Y}_k(t)
\int_{t^0}^t u_k(t', t)dt'} \label{Ysol1}
\end{eqnarray}

where we define 
\begin{eqnarray}
u_k(t', t)&\equiv& \frac{ E_k(t',t)}{\defprod ( 1 - a_{jj} \tilde{Y}_j(t) \int_{t'}^t
u_j(t'', t)dt'')^{a_{kj}/a_{jj}} } \label{usol1} \\
E_k(t', t)&\equiv& \frac{E_k(t')}{E_k(t)} \label{Ek1}
\end{eqnarray}

\noindent
Since all the $a_{kl}$ coefficients are positive, \cite{chengetal},
a system of non trivial constraints on the values of the $Y_k$'s at the scale
$t$ follows,

\begin{equation}
\tilde{Y}_k(t) < \frac{1}{a_{kk} \int_{t^0}^t u_k(t', t)dt'}
\label{landaupole}
\end{equation}

\noindent
As can be seen from Eq.(\ref{usol1}) each of the above inequalities
(for $k=1, ...n$) involves all $n$ Yukawa couplings at the same scale
$t$. The conjunction of all inequalities defines the allowed region
in the n-dimensional parameter space of $\tilde{Y}_k(t)$.
We should stress here that Eqs.(\ref{landaupole}) describe two
different, but complementary, constraints:

\noindent
- The strict inequalities
preclude the $\tilde{Y}_k$ from becoming infinite anywhere between
the low scale $t$ and the high scale
corresponding to $t^0$. Equations (\ref{landaupole}) are thus 
{\sl necessary} conditions 
guaranteeing the perturbativity  of the running Yukawa's between
$t$ and $t^0$. But they 
are obviously by no means sufficient to guarantee perturbativity; 
when going closer and closer to 
the boundary of the domain defined by Eqs.(\ref{landaupole}), higher 
order loop contributions to the running
should be included already at intermediate scales between $t$ and $t^0$, 
and a deeply non-perturbative regime is expected to set in in the vicinity of
$t^0$. {\sl Thus the LPfd conditions have a physical meaning at the one-loop 
level only when considered as {\sl necessary} perturbativity constraints.}

\noindent
- The domain defined by Eqs.(\ref{landaupole}) has yet another meaning
beyond the LPfd conditions. This is simply the positivity of the $\tilde{Y}_k$
which are by definition the squares of the Yukawa couplings appearing in the 
Lagrangian\footnote{The latter couplings are taken real throughout this paper.
Note also that for the sake of generality we avoid to consider 
Eqs.(\ref{landaupole}) as a requirement for the reality of the $u_k$
functions since such conditions are affected by the model-dependent values of
the various ratios $a_{kj}/a_{jj}$ in Eq.(\ref{usol1}).}. 
(The reader is referred to Appendix A for technical comments about
the derivation of Eqs.(\ref{landaupole}).) 

\noindent
Without the positivity constraint, but requiring a perturbativity constraint 
$|\tilde{Y}_k(t^0)| < \delta$, where $\delta$ is sufficiently small,
one can in principle allow the domain defined by the reverse inequality in
Eqs.(\ref{landaupole}). With the positivity constraint and
$\tilde{Y}_k(t^0) < \delta$ one obtains the improved bounds

\begin{equation}
\tilde{Y}_k(t) < \frac{\delta}{ u_k(t^0, t) + \delta \; a_{kk} 
\int_{t^0}^t u_k(\tau, t)d\tau}
\label{landaupole1}
\end{equation}

\noindent
which lead back to Eqs.(\ref{landaupole}) in the limit of large $\delta$.
It is customary to take $\sqrt{4 \pi}$  as the perturbative bound on the
Yukawa couplings $h_k$ at the GUT scale, that is  
$\delta \simeq 1/(4 \pi)$ when normalized as in Eq.(\ref{defeq}).    
However, one can still get back approximately Eqs.(\ref{landaupole}) when 
$a_{kk}$ is sufficiently large. This is typically what happens for the top 
quark Yukawa coupling in the standard model or the MSSM:  
the Landau pole bound and the perturbativity bound come up
numerically very close to each other even though they have theoretically 
different meanings. In the present paper we exhibit cases where this is no 
more true.

\noindent
Let us illustrate the issue with the simple model of
section $2.3$. In Fig.1 (a) we sketch the case $|a|/|b| < 1$. 
Point $B$ denotes the unique IRQFP
given by the first equation in (\ref{YQFP})  at a given
low scale $t$ in the parameter space $(\tilde{Y}_2(t), \tilde{Y}_1(t))$.
The two thick-thin dashed curves correspond each to one of the boundaries delimited by one of the two equations

\begin{eqnarray}
\tilde{Y}_1(t) &<& \frac{1}{a_{11} \int_{t^0}^t u_1(t', t)dt'} 
\label{landaupoletoy1}\\
\tilde{Y}_2(t) &<& \frac{1}{a_{22} \int_{t^0}^t u_2(t', t)dt'}
\label{landaupoletoy2}
\end{eqnarray}

\noindent
together with  (\ref{usol1}, \ref{Ek1}) wherein $k= 1, 2$
and (respectively) $j=2, 1$. The conjunction of 
Eqs.(\ref{landaupoletoy1}, \ref{landaupoletoy2})
gives the LPfd domain defined by the thick curves.
(As they stand, the above equations are highly
implicit in the variables $(\tilde{Y}_1(t), \tilde{Y}_2(t))$. 
Specific forms  within some approximations have been given elsewhere 
\cite{auberson}.)

\vspace*{1.cm}
\begin{figure}[htb]
\begin{center}
\mbox{
\begin{picture}(10,10)(0,0)
\put(199,-9){\makebox(0,0)[l]{$\tilde{Y}_2(t)$}}
\put(100,-11){\makebox(0,0)[l]{(a)}}
\put(140,160){\makebox(0,0)[l]{$\frac{|a|}{|b|} < 1$}}
\put(-8,200){\makebox(0,0)[l]{$\tilde{Y}_1(t)$}}
\put(405,-9){\makebox(0,0)[l]{$\tilde{Y}_2(t)$}}
\put(300,-11){\makebox(0,0)[l]{(b)}}
\put(320,160){\makebox(0,0)[l]{$\frac{|a|}{|b|} > 1$}}
\put(199,200){\makebox(0,0)[l]{$\tilde{Y}_1(t)$}}
\end{picture}
\psfig{figure=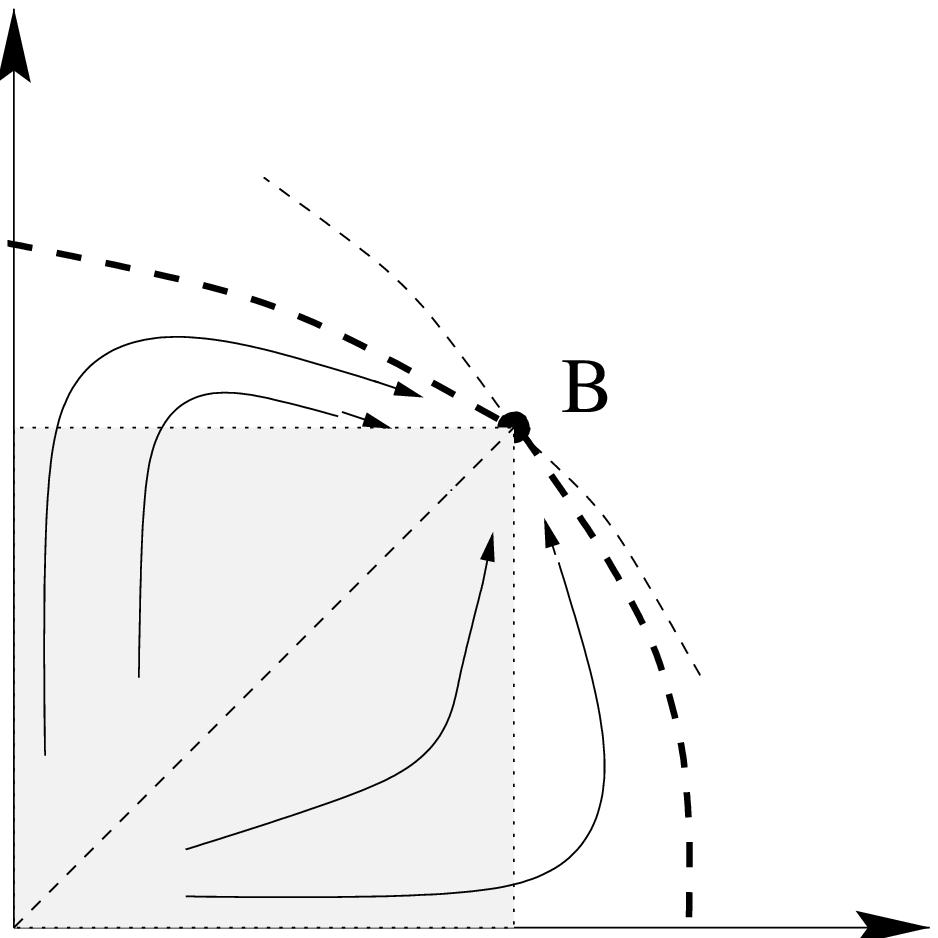,width=2.8in, height= 2.8in}

\psfig{figure=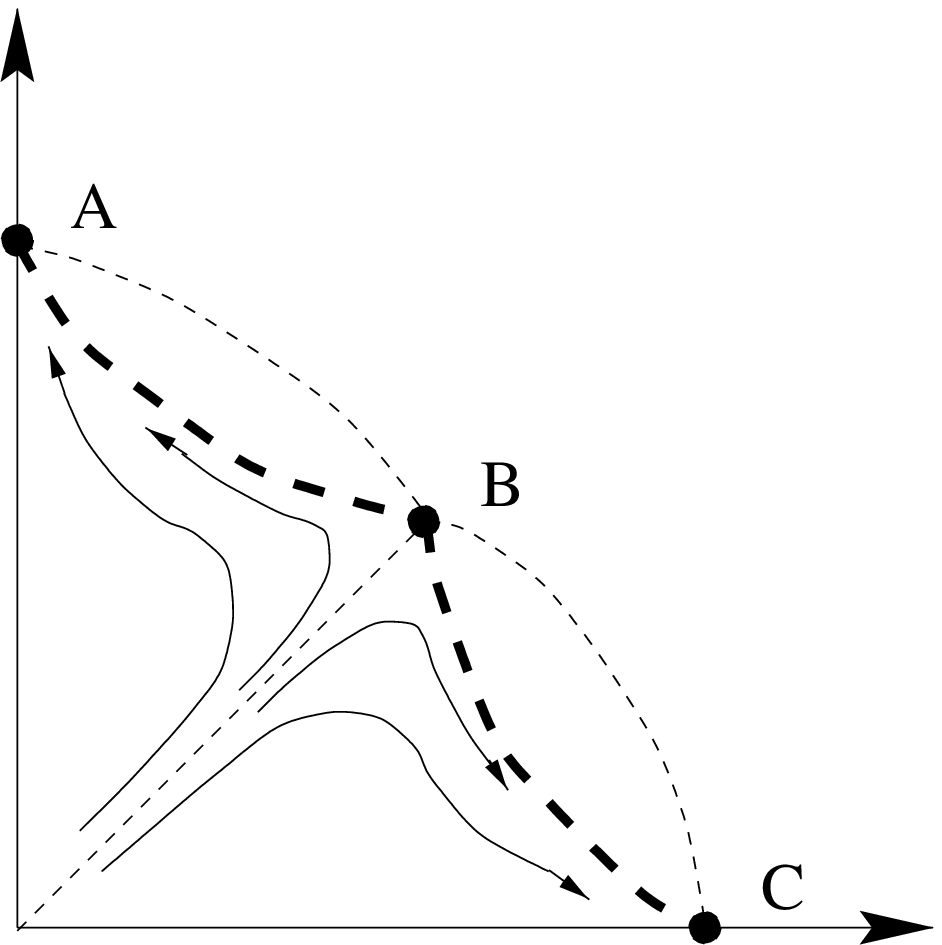,width=2.8in, height= 2.8in}}
\end{center}
\caption{ \it The IRQFP configurations for the two-Yukawa system defined
in section {\bf 2.3}. The thin lines represent the flow of 
$(\tilde{Y}_1, \tilde{Y}_2)$ at the low scale $t$, when the initial conditions 
$(\tilde{Y}_1(t^0), \tilde{Y}_2(t^0))$ are increasingly large. The different
flows correspond to a different fixed ratio between these initial conditions. } 
\end{figure}

\vspace*{1.cm}

\noindent
A point lying on one of the thick dashed lines corresponds to a configuration
where only one of the two inequalities (\ref{landaupoletoy1}, \ref{landaupoletoy2}) is saturated, that is only one of the two initial conditions 
$(\tilde{Y}_1(t^0), \tilde{Y}_2(t^0))$ is infinitely large. It is point $B$,
the unique intersection of the thick dashed lines which corresponds
to both initial conditions becoming infinite, {\sl i.e.} to the I.R. Quasi 
Fixed Point. 
This illustrates our point  {\bf (I)} in a specific context. 
The shaded rectangle would be the allowed region if one relies only
on the I.R. Quasi Fixed Point. The point here is that one is still
allowed to go {\sl above} the IRQFP value for one of the Yukawa couplings
provided that the other Yukawa coupling remains sufficiently smaller. 
Thus a complete analysis of the theoretically allowed domain and its
comparison with the eventual experimental bounds should take into account
the correlation between the Yukawa couplings. Finally we stress that
the thick dashed lines are never reached, even in the limit of infinite 
$\tilde{Y}_1(t^0), \tilde{Y}_2(t^0)$, since the only limit is point
$B$. (see appendix B.2 for further discussions). 
In Fig. 1 (a) we show typical trajectories of the Yukawa
couplings at the scale $t$. When the initial conditions at $t^0$ are
taken large (with any fixed finite ratio $\tilde{Y}_1(t^0)/\tilde{Y}_2(t^0)$), 
all the trajectories are attracted towards $B$,
even though they can still go outside of the rectangle domain. 
Although only qualitative and restrained to the case of two
Yukawa couplings, the above features are quite generic to an arbitrary
of Yukawa couplings, provided that Eq.(\ref{matricegen}) has
a unique solution. This is the case for the MSSM (including all Yukawa
couplings) \cite{kazakov} and for the (M+1)SSM \cite{mambrini}.
The next section is devoted to the intricate configuration when
Eq.(\ref{matricegen}) allows for more than one solution.

\subsection{Multiple Landau Poles}

We come now to point {\bf (II)}, first noticed 
in \cite{comment}. Here we give a general account, illustrate the 
phenomenon in the context of the simple toy model of section $2.3$, and leave 
to section 4 the discussion of the more complicated but physically interesting
case of the MSSM with R-parity violation. 
As was pointed out in section $2.2$, the equation which controls
the asymptotic low energy behaviour of the running Yukawas when their initial
conditions are large, is in general not guaranteed to have only one solution.
To start with, such a multi-solution configuration could seem mathematically 
inconsistent and contradicting the necessary unicity of the solutions of 
Eq.(\ref{yukeq})\footnote{Indeed, the running range for $t$ does not cover 
any gauge coupling Landau Poles so that the system defined by Eq.(\ref{yukeq}) 
still satisfies locally a Lipschitz condition \cite{book}, 
whence the unicity of the solution.}. 
This is actually not the case: the 
multi-solutions are due to the strict infinity of the initial conditions, and 
the unicity of the running is in fact restored for large (but finite) such 
conditions. However, the novelty is that the multi-solutions will all play a 
role here, as they will correspond either to {\sl attractive} or to 
{\sl repulsive} Quasi Fixed points. Actually, contrary to the case of
a unique solution where the (finite) ratios 
$r_{jk}=\tilde{Y}_j(t^0)/\tilde{Y}_k(t^0)$ could be varied at will with
no effect on the quasi fixed point,  
here the values of $r_{jk}$  trigger the  quasi fixed point
towards which the system will evolve. \\

\noindent
Let us illustrate this phenomenon with the system defined by Eq.(\ref{ytoy}).
As was shown in section {\bf 2.3} and Appendix B.1, there is 
potentially three different I.R.
quasi fixed points, {\sl i.e.} three solutions to Eq.(\ref{matricetoy}).
Only one of these solutions corresponds to both asymptotic powers $p_1, p_2$ 
being smaller than one, while the two other solutions have
 respectively  $p_1<1, p_2>1$ and $p_1>1, p_2<1$. From the discussion
of section $2.2$ one then expects a quasi fixed point where both 
$\tilde{Y}_1^{\mathrm{QFP}}$ and $\tilde{Y}_2^{\mathrm{QFP}}$ are non 
vanishing, and two other quasi fixed points where one (and only one) of them 
vanishes, see Eq.(\ref{YsolFP}). The question
now is which of these quasi fixed points will be effectively operating
at low energies? We will show that the quasi fixed points with one vanishing
$\tilde{Y}^{\mathrm{QFP}}$ are the effective (attractive) ones, while the 
quasi fixed point with the two non-zero $\tilde{Y}^{\mathrm{QFP}}$'s is 
repulsive. This is illustrated in Fig.1 (b), where now the two boundary lines 
defined by Eqs.(\ref{landaupoletoy1}, \ref{landaupoletoy2}) intersect 
in three distinct points corresponding to the three solutions, and the LPfd
is delimited by the thick dashed lines. 
To understand the pattern of the flow depicted in the figure, 
let us consider a point lying close to the boundary of LPfd. 
Note first that, similarly to the case where $|a|/|b| < 1$, this point cannot 
reach the LPfd boundary (when the two initial 
conditions $\tilde{Y}_1(t^0), \tilde{Y}_2(t^0)$ become infinite) 
elsewhere than on points $A$, $B$ or $C$. 
(See appendix B.2 for a more technical discussion.) In order to guess the 
dynamical behaviour close to the boundary it is useful to re-write
Eqs.(\ref{landaupoletoy1}, \ref{landaupoletoy2}) in the form

\begin{equation}
\tilde{Y}_k(t) = \frac{1 -\epsilon_k}{a_{kk} \int_{t^0}^t u_k(t', t)dt'}
\;\;\;\;\;\; (k= 1, 2)
\label{landaupoleeps}
\end{equation}

\noindent
where $\epsilon_k > 0$. This allows to write Eq.(\ref{usol1}) in the form

\begin{equation}
u_k(t', t) = \frac{ E_k(t',t) \defprod (\int_{t^0}^t
u_j(\tau, t)d\tau)^{a_{kj}/a_{jj }}}{\defprod ( \int_{t^0}^{t'}
u_j(\tau, t)d\tau   +\epsilon_j \int_{t'}^t
u_j(\tau, t)d\tau)^{a_{kj}/a_{jj}} }. \label{usoleps}
\end{equation}

\noindent
The important point here is that the denominator in the above equation\footnote{
In all models we consider, $a_{kj}/a_{jj}$ is a positive number; for the 
two-Yukawa case of section {\bf 2.3} we took $a_{kj}/a_{jj}= |a|/|b|$.}
can potentially become very small, thus $u_k(t', t)$ very large,
when $\epsilon_j \ll 1$ and simultaneously $t' \sim t^0$. 
But this is exactly the critical region we need to study:
the behaviour of $u_k(t', t)$ for $t' \sim t^0$ would determine, through
Eq.(\ref{landaupoleeps}), whether $\tilde{Y}_k(t)$ goes to zero when
$\epsilon_k \to 0$, that is whether the system is attracted towards
points $A$ or $C$, rather than $B$, when approaching the boundary of LPfd.
Because in Eq.(\ref{usoleps}) the $u_k$'s are defined iteratively in terms of 
one another , we are lead to considerations similar to those
of section {\bf 2.2} where now in Eq.(\ref{uinfty}) $Y^0$ is replaced by
$\epsilon_k$. A more sophisticated treatment leading to equations
similar to Eq.(\ref{matricegen}) is required though and is deferred to 
appendix B.2. The bottom line is as follows: 
if we consider the hierarchical regime such as 
$\epsilon_2 \ll \epsilon_1 \lsim 1$ which corresponds to a point lying very 
close to the LPfd boundary between points $B$ and $C$, then the solutions to 
Eqs.(\ref{matriceeps1},
\ref{matriceeps2}) imply unambiguously that
$\int_{t^0}^t u_1(t', t)dt' \sim \epsilon_2^{-|\Delta|}$, where $|\Delta|$
is a positive number uniquely determined by these equations. Combined with
Eq.(\ref{landaupoleeps}), this behaviour means that the system is dynamically 
attracted to point $C$. Similarly, if we consider $\epsilon_1 \ll \epsilon_2 
\lsim 1$, then the system will be attracted to point $A$. This leads altogether
to the flow pattern of Fig.1 (b). Actually, it is still possible to land on
point $B$, but only at the price of a delicate fine-tuning corresponding to the 
trajectory satisfying $\tilde{Y}_1(t) = \tilde{Y}_2(t)$, especially when
$|a|/|b|$ is much larger than one. In Fig.2 we show a purely numerical analysis
staring directly from the differential equations (\ref{ytoy}) with $a=1 
(\mbox{resp.} \; 6)$ and $b=6  (\mbox{resp.} \; 1)$. The results confirm nicely 
the above theoretical arguments. Fig.2 (a) illustrates the insensitivity
to the values of $r_{12} \equiv \tilde{Y}^0_1/\tilde{Y}^0_2$ near
the IRQFP. In contrast, one notes the high sensitivity to this ratio
in the configuration of Fig. 2 (b), due to the presence of multiple Landau 
Poles. In particular, the attraction to point
$B$ in Fig. 2 (b) occurs at the price of fine-tuning $r_{12}$ to $1$ to less
than $0.1$ per mil, otherwise the system is strongly attracted to points $A$ 
or $C$ depending on whether  $r_{12}$ is greater or smaller than one.  

\begin{figure}[htb]
\vspace*{-7cm}
\begin{center}
\mbox{
\hspace*{-2.cm}
\psfig{figure=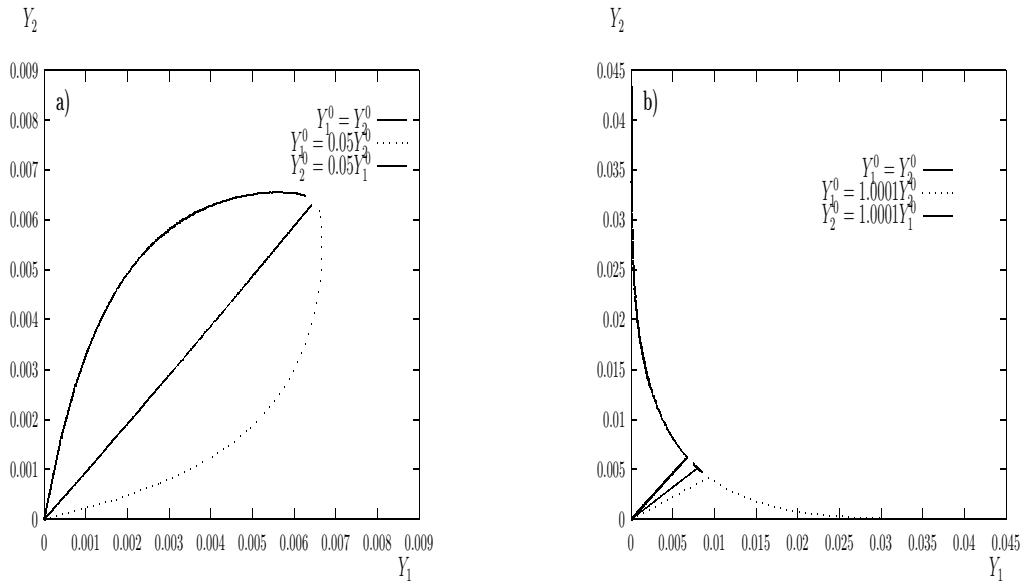, width=8.in, height= 11.in}
}
\end{center}
\vspace*{-13cm}
\caption{ \it A numerical illustration of the behaviour of the toy model 
Eqs.(\ref{ytoy}) with $|a|=1, |b|=6$ in a) and $|a|=6, |b|=1$ in b).  
The running $\tilde{Y}_1(t), \tilde{Y}_2(t)$ are evaluated at $t= 66$.
The smallest initial condition $\tilde{Y}^0_i$ is varied from 0 to 1000.
Near the IRQFP's, the insensitivity to $r_{12}$ in a) and the high sensitivity to $r_{12}$ in b) are manifest.} 
 
\end{figure}

\noindent
To summarize this section on the two-Yukawa system, we have shown in general 
that when $r_{12}$ is neither infinite nor zero, the LPfd boundary
closes only on a set of isolated points, point $B$ in the case of Fig.1 (a)
and points $A$, $B$, $C$ in the case of Fig.1 (b). Most of the LPfd boundary,
 the thick dashed lines, can be closely approached but is actually never 
reached, the system being eventually driven to one of the above points. 
In Fig.1 (a), only when $r_{12}$ is infinite or zero 
({\sl i.e.} one of the initial conditions remains finite)
does one retrieve the effective fixed point {\sl \`a la Hill}, in which case
point $B$ becomes located on one of the two axes. 
Of particular importance is the case of Fig.1 (b). It shows how the existence
of multiple IRQFP's fits nicely with the phenomenon of dynamical attraction to
or repulsion from such points, leading to the unusual suppression of some
couplings. This is quite a general mechanism which occurs
beyond the two-Yukawa system as will be illustrated in the next section.
Finally we would like to insist on the fact that although a  perturbative 
analysis is trustful only far from these IRQFP's, the latter remain nevertheless
very useful in understanding the generic behaviour of the low energy 
perturbative couplings, as they play the role of a far away beacon for such a 
behaviour.

\section{ $\not{\!\!R}_p$-MSSM}

In this section we concentrate on a phenomenological application
in the context of the MSSM with R-parity violation. The unusual behaviour
near Landau Poles was indeed first noticed in this context \cite{comment}. 

Let us first recall the form of the superpotential 
in $\not{\!\!R}_p$-MSSM:

\begin{equation}
W_{\not{R}_p} = W + W_{\not{L}} + W_{\not{B}}
\end{equation}

where $W$ is the R-parity conserving part

\begin{equation}
W= (h_E)_{i j} L_L^i.H_1 \bar{E}_R^j + (h_D)_{i j} Q_L^i.H_1 \bar{D}_R^j 
   + (h_U)_{i j} Q_L^i.H_2 \bar{U}_R^j + \mu H_1.H_2
\end{equation}

$W_{\not{L}}$  induces lepton number violation,

\begin{equation}
W_{\not{L}} = \frac{1}{2} \lambda_{ijk} L_L^i.L_L^j \bar{E}_R^k + 
                          \lambda'_{ijk} L_L^i.Q_L^j \bar{D}_R^k 
                          + \kappa_i L_L^i.H_2
\end{equation}

and $W_{\not{B}}$ induces baryon number violation,

\begin{equation}
W_{\not{B}} = \frac{1}{2} \lambda''_{ijk} \bar{U}_R^i \bar{D}_R^j \bar{D}_R^k.
\end{equation} 

\noindent
The superfields $L_L, Q_L, \bar{E}_R, \bar{D}_R, \bar{U}_R$ denote lepton and 
quark $SU(2)$ doublets and anti-singlets, $H_1,H_2$ the two Higgs doublets.
Summation on $SU(3)$ color indices is implicit and the dots 
($A \cdot B \equiv \epsilon_{a b} A_a B_b$) define SU(2) invariants
The $i,j,k=1,2,3$  are generation indices, and summation is understood
for repeated indices.  

\noindent
We will rely hereafter on \cite{allanachetal} (\cite{MartinVaughn}) 
to write down the RGE's at the one-loop level. 
The relation between the above $\lambda$'s and their
matrices ${\bf \Lambda}$  is as follows:   
$$\lambda_{ijk} \equiv ({\bf \Lambda}_{E^k})_{ij},
\lambda'_{ijk} \equiv ({\bf \Lambda}_{D^k})_{ij}, \lambda''_{ijk} \equiv ({\bf \Lambda}_{U^i})_{jk}.$$

\noindent
(Note that the notation is consistent with that of \cite{allanachetal}, but
differs for instance from \cite{pandita} only in $\lambda''$.)
The matrices ${\bf \Lambda}_{E^k}$ and 
${\bf \Lambda}_{U^k}$ being anti-symmetric, lead to a total of 18 couplings,
while ${\bf \Lambda}_{D^k}$ give 27 extra couplings. We will not consider
hereafter the bi-linear couplings $\kappa^i$ and $\mu$ as their renormalization
group evolution does not affect that of the Yukawa couplings we are
interested in, \cite{allanachetal}, \cite{MartinVaughn}.  

\noindent
The issue of theoretical bounds on RPV couplings has been addressed in
various papers \cite{bargeretal, pandita, pandita1, goity, brahma}, 
either from the point of view of {\sl exact}
infrared fixed points, when they exist, or from that of perturbativity and/or
effective fixed points related to Landau Pole considerations. However, often
a rather reduced number of couplings has been  considered which may not exhibit 
the behaviour we described in the previous sections. Typically, the larger
the number of couplings, and accordingly the dimension of ${\cal M}$ in 
Eq.(\ref{matricegen}), the more likely it is to find asymptotic powers 
significantly  greater than $1$ as solutions to this equation, leading to a 
dynamical suppression of some Yukawa couplings due to the Landau pole.

As was first shown in \cite{comment}, the system of 6 couplings 
$h_t, h_b, h_\tau, \lambda_{233}, \lambda'_{333}, \lambda''_{323}$ does
exhibit such a behaviour where $\lambda'_{333}$ decreases while the other 
couplings increase towards their perturbativity bound.\\

\noindent
On the other hand, there are also numerical studies of the simultaneous 
renormalization group evolution of all 45 Yukawa couplings, putting the accent
on the correlation between experimental bounds at low energy and initial 
values at the grand unified scale \cite{allanachetal1}. \\

\noindent
Here we take up a fairly complementary approach. We consider various
sectors with 13 couplings chosen in such a way as to encompass previous
studies as subclasses and in the same time to be realistic enough
for phenomenology. These sectors allow also a direct application of
the analytic criteria developed in the previous sections and
a clear numerical illustration of these criteria.\\

\noindent
\underline{{\bf RPV${}_{13}$}  sectors:} Each of these sectors has 13 couplings 
and is defined as follows. For fixed $k$ and fixed $l(\neq k)$ we consider the 
RGE's associated to the system

\begin{equation} 
{\bf RPV{}_{13}} \equiv ( h_t, h_b, h_\tau, \lambda_{klq}, \lambda'_{kkk}, \lambda'_{kkl}, 
\lambda'_{klk}, \lambda'_{lkk}, \lambda''_{qkl}), \;\; \mbox{where} \;\;q=1,2,3
\label{rpv13}
\end{equation}
Thus one has for each case three $\lambda$'s, three $(\lambda'')$'s and four
 $(\lambda')$'s.  The RGE's in the RPV${}_{13}$ sectors are explicitly written
in appendix C. One can see that they have strictly the form of 
Eq.(\ref{yukeq}) when $k \neq 3$. For $k=3$, Eqs.(C.2, C.3, C.6, C.10)
deviate by one extra term from this form, but as we will see throughout
the numerical analysis, the behaviour is still dictated by our
general criteria. Actually the specific choice of the set (\ref{rpv13}) is 
motivated by the opportunity to remain as close as possible to the form of 
Eqs.(\ref{yukeq}). We stress here that this is not necessarily the only 
possibility, but we stick to it as a working hypotheses which allows
for general enough, yet analytically tractable, illustrations.

\subsection{Asymptotic powers}

To start with, it is instructive to follow
the variation of the asymptotic powers 
when the number of active RPV couplings is increased.\\




{\bf RPV${}_6$:}
Consider the set of non zero couplings 
$( h_t, h_b, h_\tau, \lambda_{233}, \lambda'_{333}, 
\lambda''_{323})$. This is a subset of Eq.(\ref{rpv13}) with $k=q=3$ and
$l=2$. Here the order of the antisymmetric indices is irrelevant, since 
the evolution equations in the RPV${}_6$ subset are blind to the sign of the 
couplings, as can been seen from Appendix {\bf C}.  
In this case Eq.(\ref{matricegen}) reads 

\begin{equation}
\left(
\begin{array}{l}
p_t \cr
p_b \cr
p_{\tau} \cr
p_{233} \cr
p'_{333} \cr
p''_{323} \cr
\end{array}
\right)
=
\left(
\begin{array}{llllll}
0 & \frac{1}{6} & 0 &0 & \frac{1}{6} & \frac{1}{3}\cr
\frac{1}{6} & 0 & \frac{1}{4} & 0 & 1 &\frac{1}{3}\cr
0 & \frac{1}{2} & 0 & 1 & \frac{1}{2} &0\cr
0 & 0 &  1 & 0&  \frac{1}{2} &0 \cr
\frac{1}{6} &1& \frac{1}{4} & \frac{1}{4} & 0 & \frac{1}{3} \cr
\frac{1}{3}& \frac{1}{3}& 0 & 0 &\frac{1}{3}& 0 
\end{array}
\right)
\left(
\begin{array}{l}
(1 - p_t) \; \theta[1 - p_t] \; \delta_t \cr
(1 - p_b) \; \theta[1 - p_b] \; \delta_b \cr
(1 - p_{\tau}) \; \theta[1 - p_{\tau}] \; \delta_\tau \cr
(1 - p_{233}) \; \theta[1 - p_{233}] \; \delta_0 \cr
(1 - p'_{333}) \; \theta[1 - p'_{333}] \; \delta_1 \cr
(1 - p''_{323}) \; \theta[1 - p''_{323}] \; \delta_2 \cr
\end{array}
\right)
\label{matrice}
\end{equation}

\noindent
Solving this equation in the regime where all $\delta's = 1$, one  
finds that $ p_t, p_b, p_{233}, p''_{323} < 1$ while $p'_{333} = \frac{5}{4}$ 
and $p_\tau= \frac{23}{17}$ meaning that the running $\lambda'_{333}$ and 
$h_\tau$ at any sufficiently low energy scale will decrease to zero with 
increasing initial values at the GUT scale. However, since 
$p'_{333}, p_\tau$ are close to $1$ this decrease is relatively slow,   
\cite{comment}. Actually Eq.(\ref{matrice}) has yet another solution with all 
asymptotic powers smaller than $1$; but this solution turns out to be 
{\sl repulsive} in the sense of sections {\bf 3.2} and {\bf B.2}. In Fig. 3 we 
illustrate the evolution of the RPV${}_6$ system, solving purely numerically 
\cite{FORTRANref} the corresponding RGE's as given in appendix {\bf C}. \\

\begin{figure}[htb]
\vspace*{-5cm}
\begin{center}
\mbox{
\psfig{figure=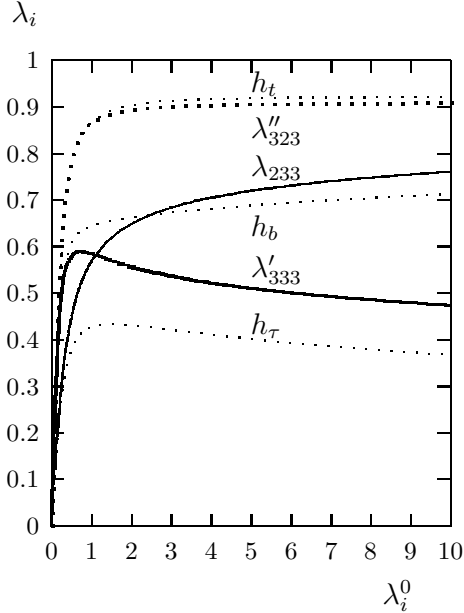, width=8.in, height= 11.in}
}
\end{center}
\vspace*{-16cm}
\caption{ \it The running couplings of the RPV${}_6$ system evaluated at 
$t=66 \simeq  Log[ M_{GUT}^2/M_Z^2]$, as a function of their common initial condition $\lambda^0$.  $\lambda'_{333}$ and $\lambda_\tau$ decrease very slowly
even for $\lambda^0 \gg \lambda^0_{pert}$.} 
\end{figure}

\noindent
The decrease of $\lambda'_{333}$ and $h_\tau$ with an increasing common value of the initial condition $\lambda^0$ nicely confirms the above expectations. 
However, since the corresponding asymptotic powers are close to $1$,
the dynamical suppression is too slow to be sizeable before the perturbativity
bound  at the GUT scale, defined for instance as 
$\lambda^0_{pert} = \sqrt{4 \pi} \simeq 3.54$, is reached.

{\bf RPV${}_{9}$:}
Adding  $\lambda'_{233}, \lambda'_{323}, \lambda'_{332}$ to the previous set of 
couplings and solving the corresponding Eq.(\ref{matricegen}) with the
help of algebraic packages \cite{MATH} 
(the relevant $9 \times 9$  matrix ${\cal M}_9$ can be straightforwardly 
extracted from ${\cal M}_{13}$ in Eq.(\ref{matrix13}) below) one now finds a 
larger value for $p'_{333} (\simeq 1.67)$. The decrease in  $\lambda'_{333}$ 
becomes strong enough to build in significantly before the perturbativity 
bound at the GUT scale is reached. Note that  there are, here too, more than
one set of solutions, actually three, but all of them have the above
value for $p'_{333}$. The dynamically chosen solution has also
$p_b \simeq  1.16$ and $p_{323} \simeq 1.3$. Again, we have illustrated
in Figs.4 (a), (b) the numerical results obtained directly from the RGE's 
of the RPV${}_9$ system and which confirm the above asymptotic power behaviour. 
Comparing Fig.3 with Fig.4 (a), we see how the activation of a bigger number
of RPV couplings strengthens the bounds on some of these couplings, 
in particular the drastic drop of $\lambda'_{333}$ in the RPV${}_9$ as
compared to the RPV${}_6$ case, or to the case of smaller number of active 
couplings like in \cite{pandita}. The price to pay is a looser bound on the 
extra couplings as shown in Fig.4 (b). 

\begin{figure}[htb]
\vspace*{-5cm}
\begin{center}
\mbox{
\hspace*{-4.cm}
\psfig{figure=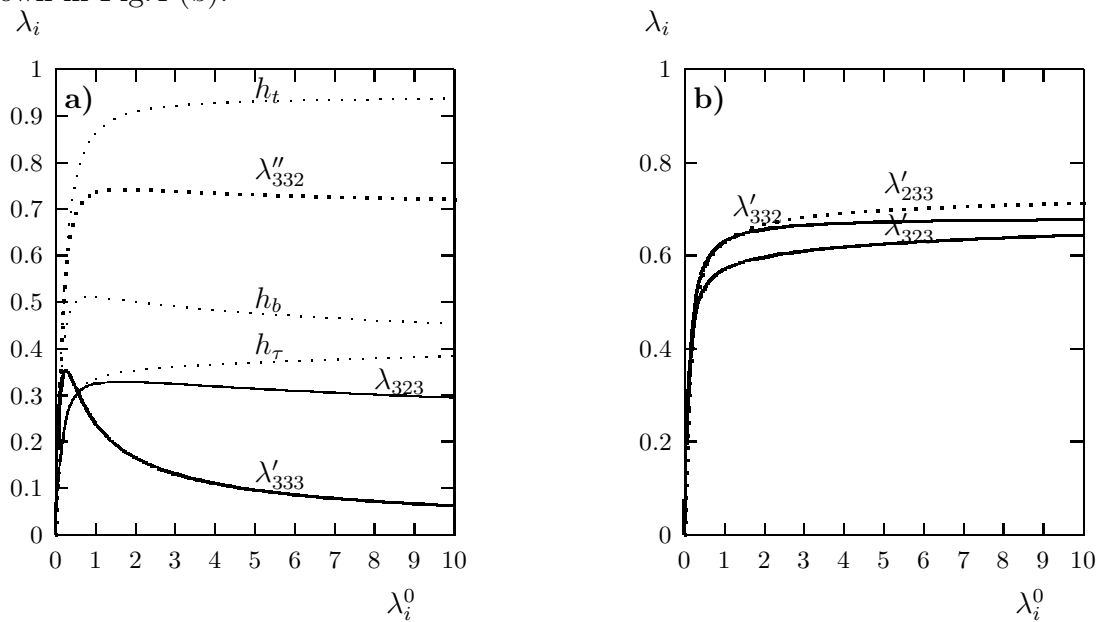, width=8.in, height= 11.in}
}
\end{center}
\vspace*{-16cm}
\caption{ \it The running couplings of the RPV${}_9$ system evaluated at 
$t=66 \simeq  Log[ M_{GUT}^2/M_Z^2]$, as a function of their common initial condition $\lambda^0$.} 
\end{figure}

{\bf RPV${}_{13}$:}
Addition of new couplings does not necessarily make the dynamical suppression
of one single coupling more pronounced, but can rather propagate this
suppression to other couplings. This is indeed the case when all 13 couplings 
as defined in Eq.(\ref{rpv13}) with $k=3$, are simultaneously activated.
The corresponding matrix of Eq.(\ref{matricedef}) is easily read from
appendix C:

\begin{equation}
{\cal M}_{13}=\left(
\begin{array}{lllllllllllll}
 0 & \frac{1}{6}& 0  &   0&   0&   0& \frac{1}{6}& \frac{1}{6}& 0 & \frac{1}{6} & 0  & 0  & \frac{1}{3} \cr
 \frac{1}{6}& 0  & \frac{1}{4}&   0&   0&   0& 1  & \frac{1}{6}& \frac{1}{3}& 1& \frac{1}{3}& \frac{1}{3}& \frac{1}{3} \cr
 0  & \frac{1}{2}& 0  & \frac{1}{4}& \frac{1}{4}&   1& \frac{1}{2}& \frac{1}{2}& \frac{1}{2}&  0& 0&   0& 0  \cr
 0  & 0  & \frac{1}{4}&  0 & 1  &   1& \frac{1}{2}& \frac{1}{2}& \frac{1}{2}&  \frac{1}{2}& 0&   0& 0  \cr
 0  & 0  & \frac{1}{4}&  1 & 0  &   1& \frac{1}{2}& \frac{1}{2}& \frac{1}{2}& \frac{1}{2}&  0&   0& 0  \cr
 0  & 0  & 1  &  1 & 1  &   0& \frac{1}{2}& \frac{1}{2}& \frac{1}{2}& \frac{1}{2}&  0&   0& 0  \cr
 \frac{1}{6}& 1  & \frac{1}{4}& \frac{1}{4}& \frac{1}{4}& \frac{1}{4}& 0  & 1  & 1  &  1  &\frac{1}{3}& \frac{1}{3}& \frac{1}{3} \cr
 \frac{1}{6}& \frac{1}{6}& \frac{1}{4}& \frac{1}{4}& \frac{1}{4}& \frac{1}{4}& 1  & 0  & \frac{1}{2}& \frac{1}{6} & \frac{1}{3}& \frac{1}{3}& \frac{1}{3} \cr
 0  & \frac{1}{3}& \frac{1}{4}& \frac{1}{4}& \frac{1}{4}& \frac{1}{4}& 1  & \frac{1}{2}& 0  & \frac{1}{3}& \frac{1}{3}& \frac{1}{3}& 
\frac{1}{3} \cr
 \frac{1}{6} & 1 & 0 & \frac{1}{4} & \frac{1}{4} & \frac{1}{4} & 1 & \frac{1}{6} & \frac{1}{3} & 0 & \frac{1}{3} & \frac{1}{3} & \frac{1}{3}  \cr 
 0  & \frac{1}{3}& 0  &0   &0   &0   & \frac{1}{3}& \frac{1}{3}& \frac{1}{3}& \frac{1}{3}& 0  & 1  & 1   \cr
 0  & \frac{1}{3}& 0  &0   &0   &0   & \frac{1}{3}& \frac{1}{3}& \frac{1}{3}& \frac{1}{3}& 1  & 0  & 1  \cr
 \frac{1}{3}& \frac{1}{3}& 0  &0   &0   &0   & \frac{1}{3}& \frac{1}{3}& \frac{1}{3}& \frac{1}{3}& 1  & 1  & 0  
\end{array}
\right) \label{matrix13}
\end{equation}

\noindent
In this case (again with all $\delta's=1$), one finds, with the
help of algebraic packages \cite{MATH}, 14 different solutions
to Eq.(\ref{matricegen}).  All these solutions have in common 
$p'_{333} \gsim 1.43$ {\sl and} $p''_{332} \sim 1.3$. In particular, the
dynamically chosen solution exhibits new features:
$p'_{333}$ has decreased slightly to $\simeq 1.46$ as compared to the RPV${}_9$ case, but now $ p_{323}$ and $p''_{332}$ are $\sim 1.3$ and 
$p'_{233} \sim 1.06$; the asymptotic powers of all other couplings remain 
smaller than one.
This means that the three couplings $\lambda_{323}, \lambda'_{333}, 
\lambda''_{332}$ are {\sl dynamically driven to zero by the Landau Pole}, 
and in any case decrease significantly, well before the perturbativity bound 
is reached, as confirmed by the numerical analysis shown in Fig.5 (a), (b); 
$\lambda'_{233}$ is also dynamically driven to zero but decreases very slowly 
since its asymptotic power is very close to $1$.

[Note that all the above conclusions hold as well for $l=1$, {\sl i.e.} for 
$\lambda_{133}, \lambda'_{333}, \lambda''_{313}$, etc... since ${\cal M}_{13}$
 is $l$ independent. ]  


\begin{figure}[htb]
\vspace*{-4.8cm}
\begin{center}
\mbox{
\hspace*{-4.cm}
\psfig{figure=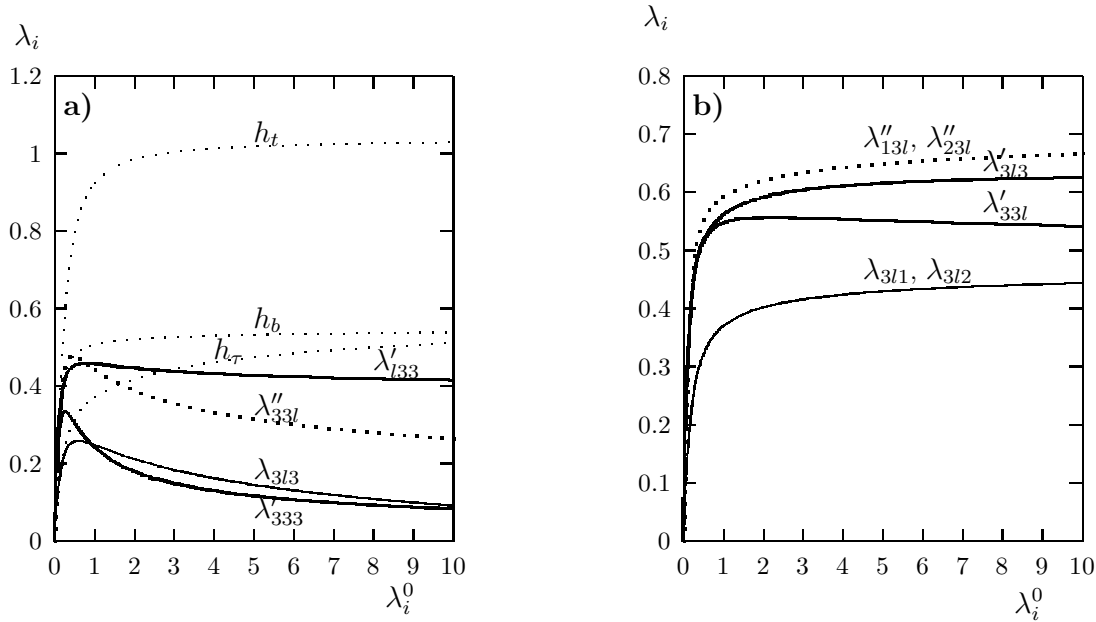, width=8.in, height= 11.in}
}
\end{center}
\vspace*{-16cm}
\caption{ \it The running couplings of the RPV${}_{13}$ system evaluated at 
$t=66 \simeq  Log[ M_{GUT}^2/M_Z^2]$, as a function of their common initial condition $\lambda^0$.} 
\end{figure}

\subsection{Theoretical bounds on the RPV couplings}
As can be seen from Figs.3,4,5, the couplings which
increase with $\lambda^0$ actually saturate and become quickly
weakly sensitive to $\lambda^0$. This is why the perturbativity
bound for the low energy couplings 
is sometimes associated to the Landau Pole. This is indeed typically the case
for the top, bottom and $\tau$ Yukawa couplings in the R-parity
conserving MSSM. However, the fact that the perturbativity bound and the
Landau Pole are in general disconnected is nicely illustrated by
those couplings which {\sl decrease} with increasing $\lambda^0$ and are 
sensitive to its values.  For instance, in RPV${}_9$ and RPV${}_{13}$,
 $\lambda'_{333}$ drops by more than a factor of 2 of its maximal
value, when $\lambda^0$ reaches $ \sqrt{4 \pi}$. Moreover, the
perturbativity bound on $\lambda'_{333}$ (taken here at $M_{susy}= 1$ TeV) 
drops 
drastically, from $0.52$ in RPV${}_6$ to $0.12 - 0.13$ in RPV${}_9$ and 
RPV${}_{13}$, as can be seen from the figures. These bounds are way below the 
ones usually quoted.\\

\noindent
In order to have a realistic comparison, also with the existing experimental 
bounds, one has to require consistency with the experimental top, bottom and 
$\tau$ fermion masses. This translates into the following constraints

\begin{equation}
h_t(m_t) = \frac{m_t \; (2 \sqrt{2} G_F)^{\frac{1}{2}} }{\sin \beta},  
\;\;\; h_{b, \tau}(m_t) = \frac{m_{b, \tau}(m_t) \;(2 \sqrt{2} G_F)^{\frac{1}{2}} }{\cos \beta} \label{fermionmass}
\end{equation}

\noindent
where $G_F$ is the Fermi constant $(=1.166 \times 10^{-5}$ GeV), $m_t$ is the 
running top quark mass at its fixed point value  $m_t(m_t) = m_t$, and 
$m_{b, \tau}(m_t)$ are the running bottom and $\tau$ masses at this value. 
We take $m_t= 165$ GeV, $m_b(m_t) = 2.76$ GeV and $m_{\tau}(m_t)= 1.78$ GeV. 
Furthermore, we will also require unification of the three gauge couplings at 
$M_{GUT}= 2 \times 10^{16}$ GeV. 
Keeping in mind that we are not seeking here a very precise gauge unification
analysis since the running is performed only at one-loop level, we tolerate
a deviation of the value of $\alpha_s(M_Z)$ from the one determined
by LEP precision measurements, as well as a susy scale $\sim M_Z$ in order
to achieve one-loop gauge unification with the proper normalization 
($ (5/3)\times g_1^2(GUT) =   g_2^2(GUT) = g_3^2(GUT)$). It should be clear, however,
that a more sophisticated treatment of this sector would not change
the main features related to the dynamical behaviour of the RPV couplings.\\

\noindent
We take the following values for the fine structure
constant $\alpha_{em}$, $\alpha_s$ and $\sin^2 \theta_W$ at the $M_Z$ scale

\begin{equation}
\alpha_{em}(M_Z)= 1/127.938, \;\; \alpha_s(M_Z) = 0.1169, \;\;
\sin^2 \theta_W(M_Z) = 0.23117
\end{equation}

\noindent
(of which  $\alpha_s$ deviates by $0.2 \%$ from the experimental value
\cite{preciseLEP}.)

\noindent
In order to study the effect of Landau Poles, the numerical algorithm will 
have to meet the  requirement of maximizing the initial Yukawa conditions at 
the GUT scale, while keeping consistency with Eq.(\ref{fermionmass}). 
This is not straightforward due to the large number of couplings. As we stressed
in previous sections, Landau Pole constraints (or for that matter,
perturbativity constraints) delimit hypervolume domains in the space
of couplings. The numerical values obtained correspond only to specific
directions in this space, depending on which set of couplings is made
large at the GUT scale. It is important to determine the directions which
allow the largest possible values for all the couplings.

In practice, we proceed as follows: for a given $\tan \beta$, and assuming a 
common susy scale much higher than the top mass, we first run the 
(non supersymmetric) standard model top, bottom and $\tau$ Yukawa couplings from
 their values at $m_t$ scale as given by Eq.(\ref{fermionmass}) up to $M_{susy}$ scale. From there we run these couplings up to the GUT scale within the 
(R-parity conserving) MSSM. There we switch on {\sl all} the RPV couplings of 
a given 
RPV${}_n$ sector, giving them in a first step a common and large value at the 
GUT scale $\lambda^{init}_{GUT}$, run the 
$n$ Yukawa couplings of the  RPV${}_n$ sector down to $M_{susy}$, re-adjust
$h_t, h_b, h_{\tau}$ to their previously determined values consistent with 
Eq.(\ref{fermionmass}) and run the RPV${}_n$ sector back to $M_{GUT}$. If at 
least one coupling is greater than the perturbative bound (taken to be 
$\sqrt{4  \pi}$) then the procedure is iterated starting from a smaller 
$\lambda^{init}_{GUT}$, in the opposite case $\lambda^{init}_{GUT}$ is 
increased and the procedure iterated until the largest coupling at the GUT 
scale reaches the limit $\sqrt{4  \pi}$ (within 1 per mil).  
However, even then, there is no 
guarantee that the other couplings have been optimally maximized, since the 
procedure starts with a unified value of the RPV's at the GUT scale. This is 
why a further step is taken after numerical convergence: start with the 
obtained solution at $M_{susy}$, scale up some of the RPV's and down some 
others, by the same factor, then run up to $M_{GUT}$ and check for 
perturbativity. Again the procedure stops after convergence to the optimal 
scaling factor. The numbers thus obtained at the susy scale correspond
to {\sl theoretical perturbativity bounds} on the various couplings. 
It is important to note here the existence of a seesaw effect between
some specific sub-classes of RPV couplings. The above inverted  scaling
works when these sub-classes are properly chosen. For instance,
in  RPV${}_{13}$, the dynamically suppressed couplings
 $\lambda_{3l3}, \lambda'_{333}, \lambda''_{33l}$ (see Fig.5) can be 
increased at the price of lowering the remaining ones.

\newpage    

\begin{figure}[htb]
\vspace*{-6cm}
\begin{center}
\mbox{
\hspace*{-4cm}
\psfig{figure=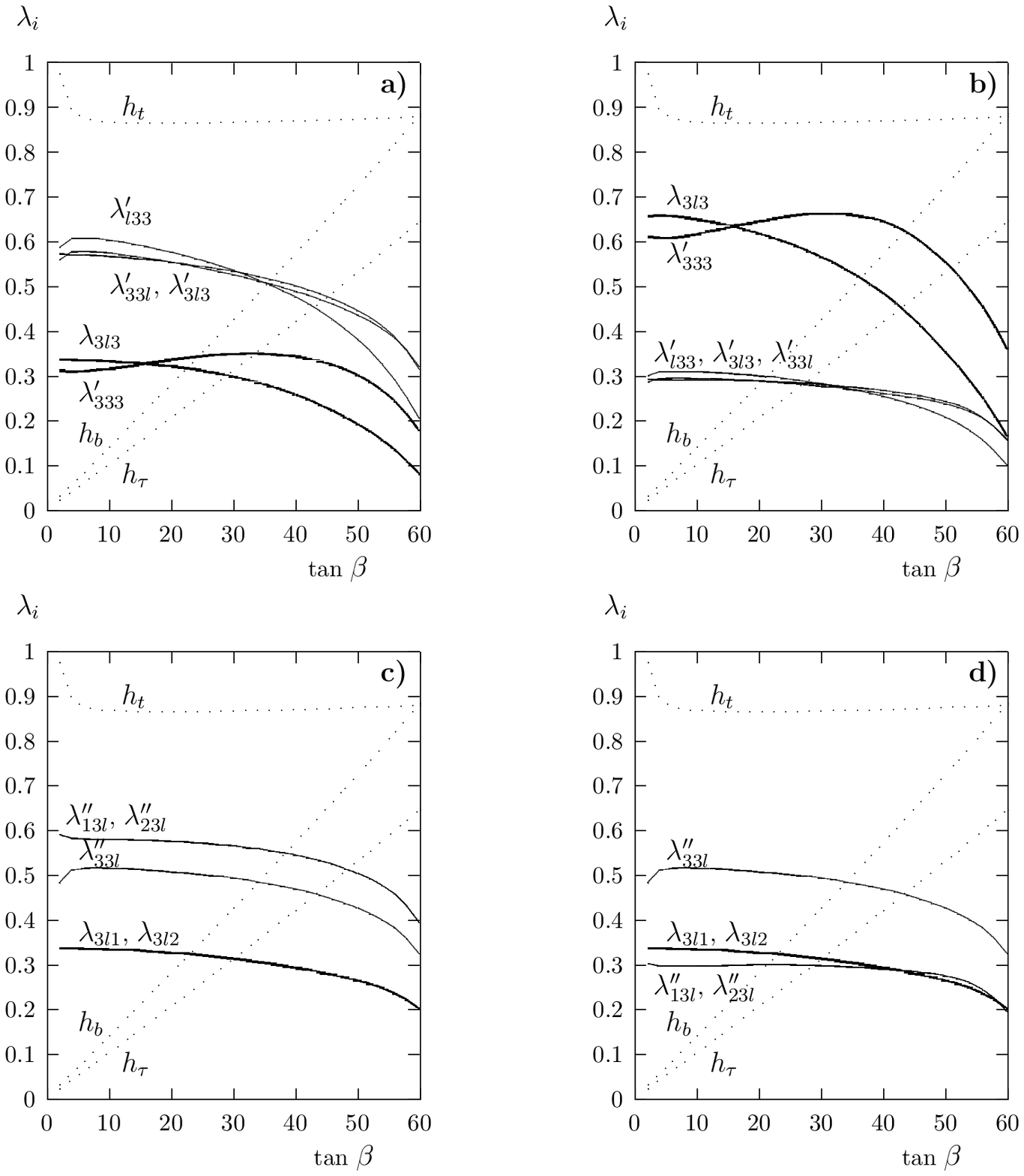,width=8.in, height= 11.in}
}
\end{center}
\vspace*{-8cm}
\caption{ \it The running couplings of the RPV${}_{13}$ system evaluated at 
the susy scale $M_{SUSY}=1$TeV ($t\simeq 61.25$), as a function of $\tan \beta$, subject to the constraints $m_t(m_t)= 165$ GeV, $m_b(m_t) = 2.76$ GeV, $m_{\tau}(m_t)= 1.78$ GeV and unification of gauge couplings at $M_{GUT}= 2 \times 10^{16}$ GeV. Figures a) and b) ( resp. c) and d) ) are two different configurations
where the perturbativity bound $\sqrt{4 \pi}$ is reached by one of the couplings
at the GUT scale.  } 
\end{figure}

\noindent
In Fig.6 we illustrate these effects as a function of $\tan \beta$ in the case
RPV${}_{13}$.
The difference between Figs. (a), (c) on one hand and
Figs. (b), (d) on the other, shows clearly that it is possible to increase,
for instance, the magnitudes of   $\lambda_{3l3}, \lambda'_{333}$ for any
value of $\tan \beta$, but various other 
$\lambda', \lambda''$ couplings should be decreased accordingly,
in order to remain consistent with the constraints described above. 
This seesaw effect is actually far more general than what is illustrated
in the figure and occurs  for the sets of $\lambda$'s, 
$(\lambda')$'s and $(\lambda'')$'s either separately or in some combinations 
of them; it is triggered by the fact that not all couplings can be
simultaneously increased (or decreased) without conflicting with
the dynamical suppression  mechanism explained before. Also the magnitude
of the effect tends to increase with decreasing $\tan \beta$. For instance,
if $\tan \beta \lsim 10$ one can raise $\lambda''_{33l}$ up to $\sim 0.97$
at the price of reducing {\sl all} the remaining RPV${}_{13}$ couplings
down to  $\lsim 0.17$ for the $\lambda$'s and $\lambda'_{333}$, and
down to $\lsim 0.3$ for the other $\lambda'$, $\lambda''$. For higher
$\tan \beta$ values these bounds drop further down: 
when $\tan \beta \sim 60$,   
$\lambda''_{33l} \sim 0.84$,  while $\lambda''_{13l}, \lambda''_{23l} \sim 0.15$, $\lambda_{3l3} \sim 3 \times 10^{-2}$, $\lambda_{3l1}, \lambda_{3l2}, 
\lambda'_{l33} \sim 7.7 \times 10^{-2}$, $\lambda'_{333} \sim 6.8 \times 
10^{-2}$ and $\lambda'_{3l3}, \lambda'_{33l} \sim 0.12 $.  
(Similar effects obtain if one chooses to raise the bound for $\lambda'_{333}$;
we will come back to that in the next section when comparing with the
present experimental limits.) We thus see that Fig.6
is with many respects a conservative illustration of the perturbativity 
bounds one can obtain.


Note that although the 
top/bottom/$\tau$ Yukawa couplings are  uniquely determined in terms of 
$\tan \beta$ through the fermion masses at the relevant 
low scale, Eq.(\ref{fermionmass}), their running values above the susy scale 
are still affected by the RPV system. In particular, the GUT scale values of 
$h_b, h_\tau$ remain small for $\tan \beta \lsim 40$ where $\lambda'_{333}$ or 
$\lambda_{3l3}$ are those which saturate the perturbativity bound, while for  
$40 \lsim \tan \beta \lsim 60$, $h_b$ then $h_\tau$ take over gradually 
in saturating this bound.\\

\noindent
The main lesson to draw from this section is the importance of including
as many couplings as possible in order to achieve a complete study
of the perturbativity bounds. The results obtained here for
the RPV${}_{13}$ system yield much stronger perturbativity bounds on 
$\lambda'$ and $\lambda''$ than previously  found  \cite{goity, brahma}. 
(see for instance the bracketed numbers displayed in the compilation 
Table 1 of ref. \cite{allanachetal1}). The latter vary in the range 
$1.04 - 1.23$ while ours can hardly reach $1$ for some couplings while 
the others are anyway bound to be much smaller. The fact that stronger bounds 
ensue from activating a larger number of couplings was already exemplified 
in \cite{goity, brahma}, since there the upper bound $1.23$ was on 
$\sqrt{ \sum_l {\lambda''_{33l}}^2}$ rather than on individual couplings. 
A similar effect is actually  observed in our numerical study, nonetheless
it should be considered as a ``phase space" effect which is complementary, 
but distinct, from the dynamical suppression effect we have studied
throughout the previous sections.


\subsection{Comparison with the experimental bounds}
Bounds on RPV couplings from experimental data have been extensively studied
in the literature, ranging from nuclear and atomic physics to astrophysics,
as well as high energy accelerator physics,
(see \cite{barbieretal}, \cite{expRPV} and references therein and there out). 
Since we are considering several $\lambda, \lambda', \lambda''$ 
in a time, a systematic comparison with the above bounds is certainly out of 
the scope of the present paper. Indeed, these bounds are usually established 
keeping a minimal set of couplings in a time \cite{barbieretal, ledroit, 
bargeretal1}, in particular for the very strong ones on the products 
$\lambda \times \lambda''$ from proton stability \cite{proton} or on $\lambda \times \lambda$, $\lambda' \times \lambda'$ from neutrino masses 
\cite{neutrino}. We shall thus limit the discussion here to
some representative examples. We stress here that the various experimental 
limits on individual couplings at the weak scale (see for instance Table 1 of 
ref. \cite{allanachetal1}), if one assumes a common susy mass of 
$\tilde{m} = 1 $ TeV, can be in general easily overwhelmed by the theoretical 
perturbativity bounds we have found in the RPV${}_{13}$ system. 
Comparing for instance to the configuration of Fig.6, the experimental limits 
on $\lambda_{321}, \lambda_{322}, \lambda_{323} \sim 0.7$ are about
 a factor of two weaker than our perturbativity bound, except for
$\lambda_{323}$ at low $\tan \beta$ in the configuration of Fig.6 (b) where
they become comparable. However, in the latter configuration the perturbativity
bounds on $\lambda'_{233}, \lambda'_{323}, \lambda'_{332} \lsim 0.3 $ are
stronger than the experimental limits\footnote{Note that even though we have 
evaluated the running couplings at $1$ TeV, these values are supposed to
be frozen and remain valid at the lower scale $M_Z$,  due to the simplifying
assumption in this particular illustration that the susy spectrum is at or 
above $m_{susy} = 1 $ TeV.}.
Of course there are configurations where the experimental limits
become, at least for some couplings, stronger than the perturbativity bounds. 
This is true in particular if one takes into account recent limits from
proton stability \cite{proton} and neutrino physics \cite{neutrino}. If
we pick up for instance the configuration where $\lambda''_{33l} \sim 0.84$
(see previous sub-section) then the perturbativity bound on the products
$\lambda_{3l3} \lambda''_{33l}, \lambda_{3l1} \lambda''_{33l}, 
\lambda_{3l2} \lambda''_{33l} \sim 2.5\times 10^{-2} - 6.5 \times 10^{-2} $ 
is overwhelmed by the proton stability limits by several orders of magnitude
\cite{proton}. For the same configuration one has the perturbativity bounds
$\lambda'_{333} \sim 6.8 \; (3.6) \; \times 10^{-2}$ and 
$\lambda_{3l3} \sim 3 \;(1.9) \;\times 10^{-2}$, where the numbers in brackets
correspond to a lower susy scale $m_{susy} = 100$ GeV. 
These bounds are clearly stronger than the experimental limits obtained
from specific $\tau$ lepton and $Z$ boson  branching ratios measurements,
but they remain two orders of magnitude weaker than those from neutrino physics
\cite{neutrino}. 
However they are still meaningful in various respects: {\sl (i)} they correspond
to a situation where many couplings are active simultaneously, while the bounds
from neutrino physics and proton stability are obtained within reduced sets or 
simplifying working assumptions \cite{neutrino, proton}, barring possible cancellations effects which
could in principle weaken substantially the latter bounds.  
{\sl (ii)} They are of the same order both for $m_{susy}=
100$ GeV or $1$ TeV, while experimental bounds from neutrino physics are 
derived assuming $m_{susy}=100$ GeV. {\sl (iii)} In both proton stability
and neutrino physics it is not clear how low energy QCD effects between
virtual quarks and squarks at $\lambda''$ and $\lambda'$ vertices 
would affect the extraction of the experimental bounds. The running RPV's
are in principle free from such uncertainties insofar as their low energy 
values are frozen at the susy scale.
  
In the case where $m_{susy} \simeq 100$ GeV, most of the $2 \sigma$ limits on 
the RPV's are a factor of ten stronger than for $m_{susy} \simeq 1$ TeV.
The perturbativity bounds are then overrun for small and moderate
$\tan \beta$ values, in particular for the $\lambda$'s of RPV${}_{13}$,
but not necessarily for the ($\lambda'$)'s and ($\lambda''$)'s. For instance,
$\tan \beta \simeq 10$ leads, for a configuration where $\lambda''_{13l}$
is maximized ($ \simeq 1.09$), to $\lambda''_{23l} \simeq 0.34, 
\lambda''_{33l} \simeq 0.28$, $\lambda'_{333} \simeq 0.18$, 
$\lambda'_{33l}, \lambda'_{3l3}, \lambda'_{l33} \simeq 0.3$, while
$\lambda_{3lq} \simeq 0.18$. For $\tan \beta$ as large as $60$, 
$\lambda''_{13l}$ maximizes to $0.62$ and $\lambda''_{23l} \simeq 0.15, \lambda''_{33l} \simeq 0.11$ 
$\lambda'_{333} \simeq 4 \times 10^{-2}$,
$\lambda'_{33l}, \lambda'_{3l3} \simeq 0.11, \lambda'_{l33} \simeq 5 
\times 10^{-2}$,  $\lambda_{3l1}, \lambda_{3l2} \simeq 7 \times 10^{-2}$ and
$\lambda_{3l3} \simeq 2 \times 10^{-2}$.  There is a tremendous drop in all
couplings which makes the above bounds stronger than most of the experimental
limits (compare Table 1 ref. \cite{allanachetal1} with $m_{susy} \sim 100$ GeV).
They remain however much weaker than the limits from neutrino physics and 
proton stability, even though some of them ({\sl e.x.} $ \lambda_{322}$) come 
very close to these strong bounds, within a factor of two.
Taking these limits at face value (albeit points {\sl (i)} -- {\sl(iii)} made 
above), they thus favour scenarios with large $\tan \beta$, which would anyway 
be what is typically needed in order to have part of the susy spectrum in the 
electroweak scale range. 

\noindent
We should keep in mind, however, that the above discussion has been carried out 
within specific configurations of the initial GUT scale conditions for the 
various RPV's, where a common value was assumed, at least as a starting point
for the numerical algorithm. Choosing hierarchical initial conditions
would allow to suppress a larger number of couplings at the electroweak scale,
as can be seen from the general structure of the solutions
Eqs.(\ref{Ysol}, \ref{uinfty}), and thus to meet eventually with stronger
experimental limits.  

To summarize, we have shown in what sense providing one single number for the 
perturbativity bound as is sometimes done, is rather restrictive and cannot
be the end of the story. Indeed, on one hand 
such a bound can vary sizeably 
with the number of active RPV couplings, and on the other hand, due to the fact
that there are perturbativity regions rather than perturbativity bounds,
much like the illustration of Fig. 1, a seesaw mechanism, involving
various RPV couplings can be operating due to the presence of attractive
and repulsive effective fixed points. 
This hints to the necessity of
including simultaneously as many RPV couplings as possible both
when determining experimental limits and when studying theoretically 
the correlations induced by renormalization group evolutions.


\section{Conclusion}

In this paper we have considered the running of an arbitrary
number of Yukawa type couplings to one-loop order, in models with an extended
sector of such couplings and a gauge sector. We studied in particular 
the general features of Landau Pole free domains for the Yukawa couplings 
and the related Infrared Quasi Fixed Points. We pinpointed the existence of 
new structures which can be interpreted in terms of multiple {\sl repulsive} 
and {\sl attractive} quasi fixed points and developed the analytical formalism 
which allows to determine such structures. An interesting consequence is the 
dynamical suppression to zero of some components of the quasi fixed points, a 
fact which contrasts with the usually expected behaviour. \\      

\noindent
We then showed that these new configurations appear naturally in the context
of the MSSM with R-parity violation. In particular, an increasing number
of R-parity violating couplings induces a suppression of some 
of these couplings when the initial conditions are large. 
A notable example is the suppression of  
$\lambda_{kl3}, \lambda'_{kkk}, \lambda''_{3kl}$ to zero at the Landau Poles.
This would be an interesting theoretical justification of the smallness of
such couplings, which goes along with stringent experimental
limits as well as the necessity of stabilizing the proton and the Neutralino 
LSP, were it not for the fact that in practice one should keep far from the 
Landau Poles for a consistent perturbative treatment. Nonetheless, taking into 
account the perturbativity bounds and constraints from the physical quark 
masses, this suppression mechanism translates into a seesaw effect involving 
all the couplings. We have demonstrated how this mechanism strengthens the  
theoretical bounds and made a quick comparison with the existing experimental 
limits. Contrary to what one could have naively expected, the simultaneous 
inclusion of a large number of R-parity couplings in the evolution equations, 
together with the limits from experimental data, can lead to more severe
bounds on these couplings than in the case where they are studied individually
or in small sets. This requires that the analyses of experimental limits
be also carried out while including simultaneously a large set of couplings. \\


{\bf Acknowledgments}\\
\noindent
We would like to thank P. Descourt and J.F. Berger for discussions and
for making available to us the lsoda fortran package,   
and are grateful to Herbi Dreiner for a useful correspondence about
the existing experimental limits. 
This work was carried out in the context of the Euro-GDR 
``supersym\'etrie". We acknowledge discussions and comments from some
of its members.

\section*{Appendix A: Technical comments on the LP conditions}
\renewcommand{\theequation}{A.\arabic{equation}}
\setcounter{equation}{0} 

It is straightforward to see from Eqs.(\ref{Ysol1}, \ref{usol1}) that  
the conditions given in Eq.(\ref{landaupole}) are {\sl sufficient} to ensure 
positivity of the $\tilde{Y}_k$'s at all scales between $t$ and $t^0$. The 
{\sl necessity} of these conditions requires some more care, since there is
still the logical possibility that some $u_k(t',t)$ and 
$1- a_{kk} \tilde{Y}_k(t) \int_{t'}^t u_k(t'', t)dt''$ may turn negative 
simultaneously  at some scale $t', t^0 \leq t' \leq t$. 
One could even imagine both those functions taking zero values simultaneously 
thus evading the Landau Pole.
We show hereafter that such a situation cannot occur. Indeed, starting
from $u_k(t,t) =1$ (see Eq. (\ref{usol1})),   
if $u_k(t',t) \to 0$ for a given $k$ and $t'$ between $t^0$ and $t$, 
then necessarily $|1- a_{jj} \tilde{Y}_j(t) \int_{t'}^t u_k(\tau, t)d\tau |
\to  \infty$ for at least one $j \neq k$ as can be seen from
Eqs.(\ref{usol1}, \ref{Ek1}, \ref{Ek}). But this can occur only 
if $|u_k(\bar{t}, t)| \to \infty$ for some $\bar{t}$ in the interval $[t', t]$.
Then in the vicinity of $\bar{t}$, the $j^{th}$ Yukawa coupling behaves
like

\begin{equation}
\tilde{Y}_j (\bar{t}) \sim \frac{u_j(\bar{t}, t)}{ -a_{jj}
\int_{\bar{t}}^t u_j(\tau, t)d\tau}
\end{equation} 

Since $a_{jj} >0$, \cite{chengetal}, and the (large) integral
is dominated by the contribution of $u_j(\bar{t}, t)$, then
$\tilde{Y}_j (\bar{t})$ has a negative sign, which contradicts its definition.
We thus conclude that no $u_k$ can vanish in the physical region
$[t^0, t]$.  

Finally, to be complete one should mention the logical possibility that
the $u_k$ functions display a non continuous jump at some scale and change 
sign without going through zero. 
In this case the positivity of the $Y_k$'s would require the denominator 
$1- a_{kk} \tilde{Y}_k(t) \int_{t'}^t u_k(\tau, t)d\tau$ 
to change sign too during the jump; since this change remains
continuous in $t'$ the denominator goes necessarily through zero.
Such a configuration exhibits a Landau Pole only at intermediate
scales between $t^0$ and $t$. Thus it does not fit to the usual
physical pattern where a Landau Pole appears only at the highest
energy scale of the problem, signaling the need for new physics 
in the vicinity of that scale. We thus disregard
this mathematical configuration altogether. 

\section*{Appendix B}

\subsection*{B.1: Non uniqueness of asymptotic behaviour}
\renewcommand{\theequation}{B.1.\arabic{equation}}
\setcounter{equation}{0}

We study the following equation,  
\begin{equation}
\left(
\begin{array}{l}
p_1 \cr
p_2 \cr
\end{array}
\right)
=
\left(
\begin{array}{cc}
\alpha & \beta \cr
\gamma & \delta \cr
\end{array}
\right)
\left(
\begin{array}{l}
(1 - p_1) \; \theta[1 - p_1] \; \delta_1 \cr
(1 - p_2) \; \theta[1 - p_2] \; \delta_2 \cr
\end{array}
\right)
\label{matrice1}
\end{equation}

\noindent
Consider the case $\delta_1=\delta_2=1$. 

{\bf (1)} If $p_1 \le 1, p_2 \le 1$ one must have

\begin{eqnarray}
p_1 &=& \frac{ \alpha + \beta + \alpha \delta - \beta \gamma}{(1 + \alpha)(1 + \delta)
                      -\beta \gamma} \le  1 \\
p_2 &=& \frac{ \gamma + \delta + \alpha \delta - \beta \gamma}{(1 + \alpha)(1 + \delta)
                      -\beta \gamma} \le  1 
\end{eqnarray}

{\bf (2)} If $p_1 \ge 1, p_2 \le 1$ one must have

\begin{eqnarray}
p_1 &=& \frac{ \beta}{1 + \delta} \ge  1 \\
p_2 &=& \frac{ \delta}{1 + \delta} \le  1 
\end{eqnarray}

{\bf (3)} If $p_1 \le 1, p_2 \ge  1$ one must have

\begin{eqnarray}
p_1 &=& \frac{ \alpha}{1 + \alpha} \le  1 \\
p_2 &=& \frac{ \gamma}{1 + \alpha} \ge  1 
\end{eqnarray}

The necessary and sufficient conditions for the simultaneous realization 
of cases {\bf (2)} and {\bf (3)}

\begin{eqnarray}
\beta \ge  1 + \delta &,& \delta > -1 \label{crit1} \\
\gamma \ge  1 + \alpha  &,& \alpha > -1 \label{crit2}
\end{eqnarray}

\noindent
are found to be necessary and sufficient also for case {\bf (1)}.
These equations are thus the criteria for the existence of three solutions
for Eq.(\ref{matrice1}). Note that there are no solutions
where $p_1, p_2$ are simultaneously $\ge 1$.

\subsection*{B.2: Attractive/repulsive IRQFP's}
\renewcommand{\theequation}{B.2.\arabic{equation}}
\setcounter{equation}{0}

Let us now consider the two-Yukawa system of section {\bf 2.3}.
In this case $\alpha=\delta=0$ and $\beta=\gamma= |a|/|b|$ in 
Eq.(\ref{matrice1}), so that 
when $|a|/|b| > 1$ there will be three different I.R.
quasi fixed points corresponding to the three configurations discussed 
in the previous subsection, 
while  $|a|/|b| < 1$ implies a unique solution associated with the 
configuration of case {\bf (1)}. Here we wish to complement, somewhat 
technically,  the qualitative discussion of the two-Yukawa system carried out 
in sections {\bf 3.1} and {\bf 3.2}. [We skip, 
though, for the sake of simplicity, the discussion of the critical
value $\frac{|a|}{|b|} = 1$.]

\noindent 
We discuss first the nature of the boundary of LPfd illustrated in
Figs.1 (a), (b). The thick dashed lines in these figures, although
representing this boundary are actually never reached, even when the two 
initial conditions are  strictly infinite, except in the points
$A$, $B$ or $C$. Indeed, in any other point on the thick dashed lines, only 
one of the two inequalities in Eqs.(\ref{landaupoletoy1}, 
\ref{landaupoletoy2}) would (by definition) saturate, implying that one and 
only one of the two initial conditions should be finite. In this case the 
structure of Eqs.(\ref{Ysol}, \ref{uinfty}) shows easily that 
(independently of the magnitudes of the asymptotic powers $p_i$) one of the two
$\tilde{Y}^{\mathrm{QFP}}$'s has to vanish, which contradicts the fact
that the point we consider was supposed to be distinct from points $A$ or $C$.

\noindent
As was discussed in section {\bf 3.2}, it is crucial to study the behaviour
of $u_k(t', t)$ as a function of $t'$ in the vicinity of $t^0$. [The scale $t$
 at which the IRQFP's are evaluated remains fixed and far from $t^0$.] Here we 
give some elements of the derivation of the equations which control this 
behaviour. The ongoing arguments are actually valid for an arbitrary number of
Yukawa couplings even though presented for simplicity in the context
of the two-Yukawa model. In view of Eq.(\ref{usoleps}), one expects
without loss of generality the following behaviour of $u_k$ in the vicinity
of $t^0$, 

\begin{equation}
u_k( t' \sim t^0, t) \sim \defprod \frac{1}{\epsilon_j^{\alpha_j}}
\end{equation}

\noindent
where the $\epsilon_j > 0 $, and the $\alpha_j$ are some real numbers
which we wish to determine. For simplicity we will assume that
there are just two types of $\epsilon_j$'s: infinitesimal ones, which we then
take all equal and denote by $\epsilon$, and non infinitesimal ones.  
Thus, depending on the regime we are interested in, taking
 $\epsilon_k = \epsilon$ for some $k$'s means that we consider a point close
 to the LPfd boundary lines defined by Eq.(\ref{landaupoleeps}) for this 
specific set of $\tilde{Y}_k$'s, but far from the others.

\noindent
In order to handle the implicit dependence on $\epsilon$ in Eq.(\ref{usoleps})
correctly, one has to split $\int_{t^0}^{t'} u_j(\tau, t)d\tau$ in the 
following way:

\begin{equation}
\int_{t^0}^{t'} u_j(\tau, t)d\tau = \int_{t^0}^{t^0 + \Delta t_1} + 
\int_{t^0 + \Delta t_1}^{t^0 + \Delta t_2} + 
\int_{t^0 + \Delta t_2}^{t^0 + \Delta t_3} + \int_{t^0 + \Delta t_3}^{t'}
\label{splitint}
\end{equation} 

\noindent
where

\begin{equation}
\Delta t_1 \ll \epsilon, \;\;\;\; \Delta t_2 \sim \epsilon, \;\;\;\; 
\epsilon \ll \Delta t_3 \ll 1
\end{equation}

\noindent
Indeed, within an expansion in the small parameter $\epsilon$ each one of these 
regions will contribute differently to the powers $\alpha_j$. Most generally 
one should then consider three different sets of $\alpha_j$, {\sl i.e.}

\begin{eqnarray}
u_k(t^0 + \Delta t, t) &\sim& \frac{1}{\epsilon^{\alpha_k^{(1)}}} \;\;\; \mbox{if}
\;\;\; \Delta t \le \Delta t_1 \label{splitregion1} \\
u_k(t^0 + \Delta t, t) &\sim& \frac{1}{\epsilon^{\alpha_k^{(2)}}} \;\;\; \mbox{if}
\;\;\;  \Delta t_1 \le \Delta t \le \Delta t_2 \label{splitregion2}\\
u_k(t^0 + \Delta t, t) &\sim& \frac{1}{\epsilon^{\alpha_k^{(3)}}} \;\;\; \mbox{if}
\;\;\;  \Delta t_2 \le \Delta t \le \Delta t_3 \label{splitregion3}
\end{eqnarray}

\noindent
Note that there is no contribution associated to the last term on the right hand
side of Eq.(\ref{splitint}) since $\tau \gg t^0$ in the corresponding 
integration region so that no $1/\epsilon$ behaviour is expected for
$u_k(\tau, t)$ there, see Eq.(\ref{usoleps}). Also one can always choose,
without loss of generality, a $\Delta t_1$ sufficiently smaller than $\epsilon$
so that the first term  on the right hand side of Eq.(\ref{splitint}) can 
always be neglected and there is no contribution from Eq.(\ref{splitregion1}). 
Using Eqs.(\ref{splitint}--\ref{splitregion3}) in Eq.(\ref{usoleps}) one 
obtains, after some straightforward but tedious analysis, a system of coupled 
equations for the $\alpha_k^{(2),(3)}$ as follows,

\begin{eqnarray}
\vec{{\cal A}}^{(2)}&=& {\cal M} \cdot \vec{ {\cal A}}^{(2)}_\theta  \label{matriceeps1} \\
\vec{{\cal A}}^{(3)}&=& {\cal M}\cdot \vec{ {\cal A}}^{(3)}_\theta \label{matriceeps2} 
\end{eqnarray}

\noindent
where the matrix ${\cal M}$ is given by Eq.(\ref{matricedef})
and $\vec{{\cal A}}, \vec{{\cal A}}_\theta$ are column vectors
defined by 

\begin{eqnarray}
(\vec{{\cal A}}^{(x)})_j &\equiv& \alpha^{(x)}_j \\
(\vec{{\cal A}}^{(x)}_\theta)_j &\equiv& \Theta^{(x)}_j \; \delta_j +   
\bar{\Theta}_j \; (1 - \delta_j), \;\;\; \;\;\; (x=2,3)
\end{eqnarray}    

\noindent
with

\begin{equation}
\Theta^{(2)}_j \equiv 1 - \theta [ \alpha_j^{(2)} - \alpha_j^{(3)}] \; 
\theta [\alpha_j^{(2)}] \; 
\alpha_j^{(2)} - \theta [ \alpha_j^{(3)} - \alpha_j^{(2)}] \;
 \theta [\alpha_j^{(3)}] \; \alpha_j^{(3)} ~~~~~~~~~~~~~~~~~~~~~~~~~~
\end{equation}
\begin{equation}
\bar{\Theta}_j \equiv \;\theta[ \alpha_j^{(2)} - \alpha_j^{(3)} - 1] \;
 \theta[\alpha_j^{(2)} - 1] \; ( 1 - \alpha_j^{(2)}) - 
\theta[ \alpha_j^{(2)} - \alpha_j^{(3)} + 1] \; 
\theta[\alpha_j^{(3)}] \; \alpha_j^{(3)}   ~~
\end{equation}
\begin{equation}
\Theta^{(3)}_j \equiv  1 - \theta [ \alpha_j^{(2)} - \alpha_j^{(3)} - 1] \;
 \theta [\alpha_j^{(2)}] \;  
\alpha_j^{(2)} - \theta [ \alpha_j^{(3)} - \alpha_j^{(2)} + 1] \;
\theta [\alpha_j^{(3)} + 1] \;(\alpha_j^{(3)} + 1 ) 
\end{equation}

\noindent
Here $\theta$ is the Heaviside function, and
$\delta_j = 1$ (resp. $0$), if $\epsilon_j = \epsilon$ (resp. $\epsilon_j \gg 
\epsilon$ ). Thus, the $\delta_j$'s parameterize the various configurations
of closeness and remoteness from the various LPfd boundary hypersurfaces. [For
instance, in the two-Yukawa case of section {\bf 3.2}, $\delta_t=1, \delta_b=0$
correspond to the hierarchical configuration $ \epsilon \equiv \epsilon_1 \ll \epsilon_2$.] The above formulation is quite general, applicable to any number
of Yukawa couplings. We actually tested it with Mathematica, up to the case of 
6 couplings, the $RPV_6$ system of section {\bf 4}. Here we stick for 
simplicity to the two-Yukawa system with $|a|/|b|> 1$. 
Solving explicitly  the coupled
equations (\ref{matriceeps1}, \ref{matriceeps2}) in the
regime $ \epsilon \equiv \epsilon_1 \ll \epsilon_2$, we find only two 
consistent sets of solutions, 
($\alpha_t^{(2)}= \alpha_t^{(3)}=-(a/b)^2, 
\alpha_b^{(2)}=\alpha_b^{(3)}=|a|/|b|$) or
 ($\alpha_t^{(2)}= \alpha_t^{(3)}=\alpha_b^{(3)}=0, \alpha_b^{(2)}=|a|/|b|$). 
Taking into account Eqs.(\ref{splitint}, \ref{splitregion2} 
\ref{splitregion3}), 
the first solution implies that in Eq.(\ref{landaupoleeps}) 
 $\int_{t^0}^t u_2 \sim \epsilon^{-|a|/|b|} + \epsilon^{1-|a|/|b|} + 
\mbox{finite part}
 \sim \epsilon^{-|a|/|b|}$ so that $\tilde{Y}_2(t) \sim \epsilon^{|a|/|b|} 
\to 0$, and similarly the second solution leads to 
$\tilde{Y}_2(t) \sim \epsilon^{|a|/|b|-1} \to 0$, while in both cases 
$\tilde{Y}_1(t)$ remains finite. This proves in general the behaviour
shown in Fig.1(b), where the flow is repelled from point $B$ and attracted
to point $A$ with a strength increasing with  $|a|/|b| ( > 1)$. 
A similar behaviour occurs for point $C$ in the regime  
$ \epsilon \equiv \epsilon_2 \ll \epsilon_1$.

\section*{Appendix C: One-loop RGE's in RPV${}_{13}$ sectors}
\renewcommand{\theequation}{C.\arabic{equation}}
\setcounter{equation}{0}

Following \cite{allanachetal}, we write here explicitly the renormalization
group equations to one-loop order which govern the evolution of the system
of 13 couplings (or less) defined in Eq.(\ref{rpv13}). 
The following RG equations are valid for one given $k$ and $l$,  $k=1,2,3$ and 
$l\neq k$. 
It is worth noting the extra terms in Eqs.(C.2, C.3, C.6, C.10) when $k=3$. 
These terms violate the general form of Eq.(\ref{yukeq}) and the insensitivity 
to the sign of the various couplings. As they stand, all couplings are assumed
to be positive. A change of sign in  one of the couplings 
$\lambda^{'}_{lkk}, \lambda_{kl3}, h_\tau, h_b$ translates into a sign flip
in front of the square roots. 
 
\begin{eqnarray}
&& \! \! 16 \pi^2 \frac{d}{dt} h_t^2 \! = \! h_t^2 \{\frac{13}{9} g_1^2 + 3 g_2^2 + 
       \frac{16}{3} g_3^2 - 
       6 h_t^2 - h_b^2 - (\lambda^{'2}_{kkk} +\lambda^{'2}_{kkl} + \lambda^{'2}_{lkk}) \delta_{k3}   
        - (1 \!-\! \delta_{k3}) \lambda^{'2}_{k3k}  - 2 \lambda^{''2}_{3kl}\}
\;\;\;\;\;\;\;\;\;\;\;\nonumber \\ &&  \\
      && \! \! 16 \pi^2 \frac{d}{dt} h_b^2\! = \!h_b^2 \{\frac{7}{9} g_1^2 + 3 g_2^2 + \frac{16}{3} g_3^2 -
       h_t^2 - 6 h_b^2 - h_\tau^2 - (6 \lambda^{'2}_{kkk} + \lambda^{'2}_{kkl} + 2 \lambda^{'2}_{klk}
       + 6 \lambda^{'2}_{lkk}    \nonumber \\
        && \! \! \;\;\;\;\;\;\;\;\;\;\;\;\;\;\;\;\;\;\;\;\;\;\;\; + 2 (\lambda^{''2}_{1kl} + \lambda^{''2}_{2kl} + \lambda^{''2}_{3kl} ) )\delta_{k3}
       - ( 1 - \delta_{k3}) (\lambda^{'2}_{k3k} + 2 \lambda^{'2}_{kk3} + 2 (\lambda^{''2}_{1k3} + \lambda^{''2}_{2k3} + \lambda^{''2}_{3k3} ) ) \}
       \nonumber \\
       &&\! \! \;\;\;\;\;\;\;\;\;\;\;\;\;\;\;\;\;\;\;\;\;\;\;\;- \sqrt{\lambda^{'2}_{lkk} \lambda^2_{kl3} h_\tau^2 h_b^2} \delta_{k3}\\ && \nonumber \\ 
      && \! \! 16 \pi^2 \frac{d}{dt} h_\tau^2\! = \!h_\tau^2 \{3 g_1^2+3 g_2^2-3 h_b^2-4 h_\tau^2
      - (\lambda^2_{kl1} +\lambda^2_{kl2} +4 \lambda^2_{kl3} +3 \lambda^{'2}_{kkk} +3 \lambda^{'2}_{kkl} + 3 \lambda^{'2}_{klk}) \delta_{k3}
       \nonumber \\ 
       &&  \! \! \;\;\;\;\;\;\;\;\;\;\;\;\;\;\;\;\;\;\;\;\;\;\;\;- (1 - \delta_{k3})(  \lambda^{2}_{k31} + \lambda^{2}_{k32} + 4 \lambda^{2}_{k33})\}
       - 3 \sqrt{\lambda^{'2}_{lkk} \lambda^2_{kl3} h_\tau^2 h_b^2} \delta_{k3}
\end{eqnarray}

\begin{eqnarray}
      && \! \! 16 \pi^2 \frac{d}{dt} \lambda^2_{kl1}\! = \!\lambda^2_{kl1} \{3 g_1^2+3 g_2^2 -h_\tau^2 (\delta_{k3} + \delta_{l3}) 
        -4 \lambda^2_{kl1}-4 \lambda^2_{kl2} 
       -4 \lambda^2_{kl3} -3 (\lambda^{'2}_{kkk} +\lambda^{'2}_{kkl} + \lambda^{'2}_{klk} + \lambda^{'2}_{lkk}) \} \;\;\;\;\;\;\;\nonumber \\ &&\\ 
      && \! \! 16 \pi^2 \frac{d}{dt} \lambda^2_{kl2}\! = \!\lambda^2_{kl2} \{3 g_1^2+3 g_2^2 -h_\tau^2 (\delta_{k3} + \delta_{l3})
        -4 \lambda^2_{kl1}-4 \lambda^2_{kl2} 
       -4 \lambda^2_{kl3} -3 (\lambda^{'2}_{kkk} +\lambda^{'2}_{kkl} + \lambda^{'2}_{klk} + \lambda^{'2}_{lkk}) \}\nonumber \\ && \\ 
      && \! \! 16 \pi^2 \frac{d}{dt} \lambda^2_{kl3} \! = \!\lambda^2_{kl3} \{3 g_1^2+3 g_2^2 -2 (1 + \delta_{k3} + \delta_{l3}) h_\tau^2 
         -4 \lambda^2_{kl1}-4 \lambda^2_{kl2}  -4 \lambda^2_{kl3}  \nonumber \\
&& \! \! \;\;\;\;\;\;\;\;\;\;\;\;\;\;\;\;\;\;\;\;\;\;\;\; \;\;\;\;\;\;\;\;\;\;\;\;\;\;\;\;\;\;\;\;\;\;\;\;\;\;\
-3 (\lambda^{'2}_{kkk} 
            +\lambda^{'2}_{kkl} + \lambda^{'2}_{klk} + \lambda^{'2}_{lkk}) \}
-3 \sqrt{\lambda^{'2}_{lkk} \lambda^2_{kl3} h_\tau^2 h_b^2} \delta_{k3}
\end{eqnarray}

\begin{eqnarray}
        && \! \! 16 \pi^2 \frac{d}{dt} \lambda^{'2}_{kkk} \! = \!\lambda^{'2}_{kkk} \{\frac{7}{9} g_1^2+3 g_2^2+\frac{16}{3} g_3^2 
      - (h_t^2+6 h_b^2+h_\tau^2) \delta_{k3}  -\lambda^2_{kl1} - \lambda^2_{kl2} - \lambda^2_{kl3} \nonumber \\
       && \! \! \;\;\;\;\;\;\;\;\;\;\;\;\;\;\;\;\;\;\;\;\;\;\;\;\;\;\;\;- 6 (\lambda^{'2}_{kkk} + \lambda^{'2}_{kkl} + \lambda^{'2}_{klk} + \lambda^{'2}_{lkk}) 
       -2 ( \lambda^{''2}_{1kl} + \lambda^{''2}_{2kl} + \lambda^{''2}_{3kl} ) \}\\ && \nonumber \\
      && \! \! 16 \pi^2 \frac{d}{dt} \lambda^{'2}_{kkl} \! = \!\lambda^{'2}_{kkl} \{\frac{7}{9} g_1^2+3 g_2^2+\frac{16}{3} g_3^2 
       - (h_t^2 + h_b^2 + h_\tau^2) \delta_{k3} - 2 h_b^2 \delta_{l3}  - \lambda^2_{kl1} - \lambda^2_{kl2} - \lambda^2_{kl3} \;\;\;\;\;\;\;\;\;\;\;\;\;\;\;\;\;\;\;\;\;\;\;\;\;
           \nonumber \\
       && \! \! \;\;\;\;\;\;\;\;\;\;\;\;\;\;\;\;\;\;\;\;\;\;\;\;\;\;\;\; 
- 6 (\lambda^{'2}_{kkk} + \lambda^{'2}_{kkl}) - 3 \lambda^{'2}_{klk} - \lambda^{'2}_{lkk} 
       -2 ( \lambda^{''2}_{1kl} + \lambda^{''2}_{2kl} + \lambda^{''2}_{3kl} ) \}\\ && \nonumber \\
      && \! \! 16 \pi^2 \frac{d}{dt} \lambda^{'2}_{klk} \! = \!\lambda^{'2}_{klk} \{\frac{7}{9} g_1^2+3 g_2^2+\frac{16}{3} g_3^2 
       - (2 h_b^2 + h_\tau^2) \delta_{k3}  - (h_t^2 + h_b^2) \delta_{l3}- \lambda^2_{kl1} - \lambda^2_{kl2} - \lambda^2_{kl3} 
            \nonumber \\
       && \! \! \;\;\;\;\;\;\;\;\;\;\;\;\;\;\;\;\;\;\;\;\;\;\;\;\;\;\;\; 
- 6 \lambda^{'2}_{kkk} -3 \lambda^{'2}_{kkl}
- 6 \lambda^{'2}_{klk} - 2 \lambda^{'2}_{lkk} 
       -2 ( \lambda^{''2}_{1kl} + \lambda^{''2}_{2kl} + \lambda^{''2}_{3kl} ) \}\\ && \nonumber \\
      && \! \! 16 \pi^2 \frac{d}{dt} \lambda^{'2}_{lkk} \! = \!\lambda^{'2}_{lkk} \{\frac{7}{9} g_1^2+3 g_2^2+\frac{16}{3} g_3^2 
        - (h_t^2 + 6 h_b^2) \delta_{k3} -h_\tau^2 \delta_{l3} - \lambda^2_{kl1} - \lambda^2_{kl2} - \lambda^2_{kl3} \nonumber \\
       && \! \! \;\;\;\;\;\;\;\;\;\;\;\;\;\;\;\;\;\;\;\;\;\;\;\;\;\;\;\;
- 6 \lambda^{'2}_{kkk} 
       - \lambda^{'2}_{kkl}
       - 2 \lambda^{'2}_{klk} - 6 \lambda^{'2}_{lkk} 
       -2 ( \lambda^{''2}_{1kl} + \lambda^{''2}_{2kl} + \lambda^{''2}_{3kl} ) \}
       -\sqrt{\lambda^{'2}_{lkk} \lambda^2_{kl3} h_\tau^2 h_b^2} \delta_{k3}
\end{eqnarray}

\begin{eqnarray} 
      && \! \! 16 \pi^2 \frac{d}{dt} \lambda^{''2}_{1kl} \! = \!\lambda^{''2}_{1kl} \{\frac{4}{3} g_1^2+8 g_3^2 
       - 2 h_b^2 (\delta_{k3} + \delta_{l3})- 2 \lambda^{'2}_{kkk} - 2 \lambda^{'2}_{kkl} - 2 \lambda^{'2}_{klk} -2 \lambda^{'2}_{lkk} \;\;\;\;\;\;\;\;\;\;\;\;\;\;\;\;\;\;\;\;\;\;\;\;\;\;\;\;\;\;\;\;\;\;\;\;\;\;\;\nonumber \\ && \! \! \;\;\;\;\;\;\;\;\;\;\;\;\;\;\;\;\;\;\;\;\;\;\;\;\;\;\;\;\;\;\;\;\;\;\;\;\;\;\;\;\;\;\;\;\;\;\;\;\;\;\;\;\;\;\;\;\;\;\;\;\;\;\;\;\;\;\;\;\;\;\;\;\;\;\;\;\;\;\;\;\;\;\;\;\;\;\;\;\;\;\;\;\;\;\;\;\;\;
         -6 ( \lambda^{''2}_{1kl} + \lambda^{''2}_{2kl} + \lambda^{''2}_{3kl}) \}  \\ &&  \nonumber \\
      && \! \! 16 \pi^2 \frac{d}{dt} \lambda^{''2}_{2kl} \! = \!\lambda^{''2}_{2kl} \{\frac{4}{3} g_1^2+8 g_3^2 
       - 2 h_b^2 (\delta_{k3} + \delta_{l3})- 2 \lambda^{'2}_{kkk} - 2 \lambda^{'2}_{kkl} - 2 \lambda^{'2}_{klk} -2 \lambda^{'2}_{lkk} \nonumber \\ && \! \! \;\;\;\;\;\;\;\;\;\;\;\;\;\;\;\;\;\;\;\;\;\;\;\;\;\;\;\;\;\;\;\;\;\;\;\;\;\;\;\;\;\;\;\;\;\;\;\;\;\;\;\;\;\;\;\;\;\;\;\;\;\;\;\;\;\;\;\;\;\;\;\;\;\;\;\;\;\;\;\;\;\;\;\;\;\;\;\;\;\;\;\;\;\;\;\;\;\;
         -6 ( \lambda^{''2}_{1kl} + \lambda^{''2}_{2kl} + \lambda^{''2}_{3kl} ) \} \\ &&   \nonumber \\
      && \! \! 16 \pi^2 \frac{d}{dt} \lambda^{''2}_{3kl} \! = \!\lambda^{''2}_{3kl} \{\frac{4}{3} g_1^2+8 g_3^2 
       -2 h_t^2 -2 h_b^2 (\delta_{k3} + \delta_{l3})-2 \lambda^{'2}_{kkk} -2 \lambda^{'2}_{kkl} - 2 \lambda^{'2}_{klk} -2 \lambda^{'2}_{lkk} 
       \nonumber \\ && \! \! \;\;\;\;\;\;\;\;\;\;\;\;\;\;\;\;\;\;\;\;\;\;\;\;\;\;\;\;\;\;\;\;\;\;\;\;\;\;\;\;\;\;\;\;\;\;\;\;\;\;\;\;\;\;\;\;\;\;\;\;\;\;\;\;\;\;\;\;\;\;\;\;\;\;\;\;\;\;\;\;\;\;\;\;\;\;\;\;\;\;\;\;\;\;\;\;\;\;-6 ( \lambda^{''2}_{1kl} + \lambda^{''2}_{2kl} + \lambda^{''2}_{3kl} ) \} 
\end{eqnarray}



 


\newpage



\newpage

\end{document}